\documentclass[11pt,draftcls,onecolumn,peerreview,oneside]{IEEEtran}

\usepackage[english]{babel}
\usepackage[applemac]{inputenc}
\usepackage[T1]{fontenc}
\usepackage{epsfig}
\usepackage{amsfonts,amssymb,amsmath,amsthm,bm}
\usepackage{algorithmic}
\usepackage{algorithm}
\usepackage{url}
\usepackage{subfig,graphicx}
\usepackage[]{hyperref}
\usepackage{setspace}
\usepackage{rotating}
\usepackage{longtable}
\usepackage{lscape}
\usepackage{color}

\newcommand{\Vector}[1]{\bm{#1}} 						
\newcommand{\Matrix}[1]{\bm{#1}} 						
\newcommand{\Set}[1]{\mathbb{#1}} 					
\newcommand{\Function}[1]{\mathcal{#1}}	 			
\newcommand{\Label}[1]{\mathrm{#1}} 					

\newcommand{\Acronym}[1]{\textsc{#1}} 				

\newcommand{\Transpose}[1]{{#1}^{T}}						 

\newcommand{\abs}[1]{\left|#1\right|} 					
\newcommand{\Fourier}[1]{\hat{#1}} 					    

\newcommand{\optimal}[1]{#1^{\star}} 					
\newcommand{\Min}[1]{\underset{#1}{\arg \min}\;}		
\newcommand{\Minimise}[2]{\Min{#1} #2} 				

\newcommand{\roundb}[1]{\left(#1\right)} 					
\newcommand{\curlyb}[1]{\left\{#1\right\}} 				

\def \real{\Set{R}} 	
\def \sig{\Vector{s}} 	
\def \iSig{m}			
\def \nSig{M} 			
\def \cat{c} 			
\def \iCat{\iSig} 		
\def \nCat{\nSig} 		
\def \new{\Label{new}} 	
\def \iFra{n} 		  	
\def \nFraDim{D} 		
\def \tra{\Function{T}} 
\def \fea{\Vector{\feaCoeff}} 	
\def \feaCoeff{x} 		
\def \nFeaDim{K} 		
\def \iFeaDim{k} 		
\def \uniCat{\gamma} 	
\def \iUniCat{q} 		
\def \nUniCat{Q} 			
\def \idxSet{\Lambda} 		
\def \model{\Function{M}} 	
\def \lea{\Function{S}} 	
\def \dec{\Function{G}} 	
\def \crit{g} 				
\def \nMelBands{R} 			
\def \Normal{\mathcal{N}} 	
\def \mean{\Vector{\mu}} 	
\def \covar{\Matrix{\Sigma}}
\def \iCom{i} 				
\def \nCom{I} 				
\def \GaussWeight{w} 		
\def \prob{p} 				
\def \iTim{t} 				
\def \iLag{l} 				
\def \nLags{L} 				
\def \lpcCoe{\alpha}		
\def \ato{\Vector{\phi}} 	
\def \iAto{j} 				
\def \nAtos{J} 				
\def \parTra{\theta} 		
\def \iTes{p} 				
\def \nTes{P} 				
\def \Classifier{\Function{C}} 
\def \acc{\Gamma} 			
\def \BernoulliVar{X} 		
\def \SignVar{Y} 			
\def \cumAcc{\rho} 			
\begin{document}

\title{Acoustic Scene Classification}

\author{
	\IEEEauthorblockN{Daniele~Barchiesi, Dimitrios Giannoulis, Dan Stowell and Mark~D.~Plumbley,~\IEEEmembership{Senior Member,~IEEE.}}%
	\IEEEauthorblockA{
		School of Electronic Engineering and Computer Science\\
		Queen Mary University of London\\
		Mile End Road, London E1 4NS, UK\\
		Email: <firstname.secondname>@qmul.ac.uk\\
		}}
\thanks{Copyright (c) 2012 IEEE. Personal use of this material is permitted. However, permission to use this material for any other purposes must be obtained from the IEEE by sending a request to pubs-permissions@ieee.org.}
\thanks{This work was supported by the Centre for Digital Music Platform Grant EP/K009559/1 and a Leadership Fellowship EP/G007144/1 both from the UK Engineering and Physical Sciences Research Council (EPSRC).}%

\maketitle

\begin{abstract}
In this article we present an account of the state-of-the-art in acoustic scene classification (\Acronym{ASC}), the task of classifying environments from the sounds they produce. Starting from a historical review of previous research in this area, we define a general framework for \Acronym{ASC} and present different implementations of its components. We then describe a range of different algorithms submitted for a data challenge that was held to provide a general and fair benchmark for \Acronym{ASC} techniques. The dataset recorded for this purpose is presented, along with the performance metrics that are used to evaluate the algorithms and statistical significance tests to compare the submitted methods. We use a baseline method that employes \Acronym{MFCCs}, \Acronym{GMMs} and a maximum likelihood criterion as a benchmark, and only find sufficient evidence to conclude that three algorithms significantly outperform it. We also evaluate the human classification accuracy in performing a similar classification task. The best performing algorithm achieves a mean accuracy that matches the median accuracy obtained by humans, and common pairs of classes are misclassified by both computers and humans. However, all acoustic scenes are correctly classified by at least some individuals, while there are scenes that are misclassified by all algorithms.
\end{abstract}

\section{Introduction}
Enabling devices to make sense of their environment through the analysis of sounds is the main objective of research in \emph{machine listening}, a broad investigation area related to computational auditory scene analysis (\Acronym{CASA})\cite{wang2006}. Machine listening systems perform analogous processing tasks to the human auditory system, and are part of a wider research theme linking fields such as machine learning, robotics and artificial intelligence.

Acoustic scene classification (\Acronym{ASC}) refers to the task of associating a semantic label to an audio stream that identifies the environment in which it has been produced. Throughout the literature on \Acronym{ASC}, a distinction is made between psychoacoustic/psychological studies aimed at understanding the human cognitive processes that enable our understanding of acoustic scenes  \cite{McAdams1993Re}, and computational algorithms that attempt to automatically perform this task using signal processing and machine learning methods. The perceptual studies have also been referred to as soundscape cognition \cite{Dubois2006A-}, by defining \emph{soundscapes} as the auditory equivalent of landscapes \cite{Schafer1977Th}. In contrast, the computational research has also been called computational auditory scene recognition \cite{Peltonen2002CO}. This is a particular task \textcolor{red}{that is related to the area} of \Acronym{CASA} \cite{wang2006}, and is especially applied to the study of environmental sounds \cite{Ellis1996Pr}. \textcolor{red}{It is worth noting that, although many \Acronym{ASC} studies are inspired by biological processes, \Acronym{ASC} algorithms do not necessarily employ frameworks developed within \Acronym{CASA}, and the two research fields do not completely overlap}. In this paper we will mainly focus on computational research, though we will also present results obtained from human listening tests for comparison. 

Work in \Acronym{ASC} has evolved in parallel with several related research problems. For example, methods for the classification of noise sources have been employed for noise monitoring systems \cite{Gaunard1998AU} or to enhance the performance of speech-processing algorithms \cite{El-Maleh1999Fr}. Algorithms for sound source recognition \cite{Defreville2006Au} attempt to identify the sources of acoustic events in a recording, and are closely related to event detection and classification techniques. The latter methods are aimed at identifying and labelling temporal regions containing single events of a specific class and have been employed, for example, in surveillance systems \cite{Radhakrishnan2005AU}, elderly assistance \cite{Guyot2013Wa} \textcolor{red}{and speech analysis through segmentation of acoustic scenes \cite{hu2007auditory}}. Furthermore, algorithms for the semantic analysis of audio streams that also rely on the recognition or clustering of sound events have been used for personal archiving \cite{Ellis2004Mi} and audio segmentation \cite{Lie-Lu2002Co} and retrieval \cite{Zhang2001Au}. 

The distinction between event detection and \Acronym{ASC} can sometimes appear blurred, for example when considering systems for multimedia indexing and retrieval \cite{Chaudhuri2011Un} where the identification of events such as the sound produced by a baseball hitter batting in a run also characterises the general environment \emph{baseball match}. On the other hand, \Acronym{ASC} can be employed to enhance the performance of sound event detection \cite{Heittola2013Co} by providing prior information about the probability of certain events. To limit the scope of this paper, we will only detail systems aimed at modelling complex physical environments containing multiple events.

Applications that can specifically benefit from \Acronym{ASC} include the design of context-aware services \cite{Schilit1994Co}, intelligent wearable devices \cite{Xu2008In}, robotics navigation systems \cite{Chu2006WH} and audio archive management \cite{Landone2007En}. Concrete examples of possible future technologies that could be enabled by \Acronym{ASC} include smartphones that continuously sense their surroundings, switching their mode to silent every time we enter a concert hall; assistive technologies such as hearing aids or robotic wheelchairs that adjust their functioning based on the recognition of indoor or outdoor environments; or sound archives that automatically assign metadata to audio files. Moreover, classification could be performed as a preprocessing step to inform algorithms developed for other applications, such as source separation of speech signals from different types of background noise. Although this paper details methods for the analysis of audio signals, it is worth mentioning that to address the above problems acoustic data can be combined with other sources of information such as geo-location, acceleration sensors, collaborative tagging and filtering.

From a purely scientific point of view, \Acronym{ASC} represents an interesting problem \textcolor{red}{that both humans and machines are only able to solve to a certain extent}. From the outset, semantic labelling of an acoustic scene or soundscape is a task open to different interpretations, as there is not a comprehensive taxonomy encompassing all the possible categories of environments. Researchers generally define a set of categories, record samples from these environments, and treat \Acronym{ASC} as a supervised classification problem within a closed universe of possible classes. Furthermore, even within pre-defined categories, the set of acoustic events or qualities characterising a certain environment is generally unbounded, making it difficult to derive rules that unambiguously map acoustic events or features to scenes.

In this paper we offer a tutorial and a survey of the state-of-the-art in \Acronym{ASC}. We provide an overview of existing systems, and a framework that can be used to describe their basic components. We evaluate different techniques using signals and performance metrics especially created for an \Acronym{ASC} signal processing challenge, and compare algorithmic results to human performance.


\section{Background: a history of acoustic scene classification}\label{sec:back}
The first method appearing in the literature to specifically address the \Acronym{ASC} problem was proposed by Sawhney and Maes  \cite{Sawhney1997Si} in a 1997 technical report from the \Acronym{MIT} Media Lab. The authors recorded a dataset from a set of classes including `people', `voices', `subway', `traffic', and `other'. They extracted several features from the audio data using tools borrowed from speech analysis 
and auditory research
, employing recurrent neural networks 
and a k-nearest neighbour criterion to model the mapping between features and categories, and obtaining an overall classification accuracy of $68\%$. A year later, researchers from the same institution \cite{Clarkson1998Au} recorded a continuous audio stream by wearing a microphone while making a few bicycle trips to a supermarket, and then automatically segmented the audio into different scenes (such as `home', `street' and `supermarket'). For the classification, they fitted the empirical distribution of features extracted from the audio stream to Hidden Markov Models (\Acronym{HMM}). 

Meanwhile, research in experimental psychology was focussing on understanding the perceptual processes driving the human ability to categorise and recognise sounds and soundscapes. Ballas \cite{Ballas1993Co} found that the speed and accuracy in the recognition of sound events is related to the acoustic nature of the stimuli, how often they occur, and whether they can be associated with a physical cause or a sound stereotype. Peltonen \emph{et. al.} \cite{Peltonen2001Re} observed that the human recognition of soundscapes is guided by the identification of typical sound events such as human voices or car engine noises, and measured an overall $70\%$ accuracy in the human ability to discern among $25$ acoustic scenes. Dubois \emph{et al.} \cite{Dubois2006A-} investigated how individuals define their own taxonomy of semantic categories when this is not given a-priori by the experimenter. Finally, Tardieu \emph{et al.} \cite{Tardieu2008Pe} tested both the emergence of semantic classes and the recognition of acoustic scenes within the context of rail stations. They reported that sound sources, human activities and room effects such as reverberation are the elements driving the formation of soundscape classes and the cues employed for recognition when the categories are fixed a-priori.

Influenced by the psychoacoustic/psychological literature that emphasised both local and global characteristics for the recognition of soundscapes, some of the computational systems that built on the early works by researchers at the \Acronym{MIT} \cite{Sawhney1997Si,Clarkson1998Au} focussed on modelling the temporal evolution of audio features. Eronen \emph{et al.} \cite{Eronen2003Au} employed Mel-frequency cepstral coefficients (\Acronym{MFCCs}) 
to describe the local spectral envelope of audio signals, and Gaussian mixture models (\Acronym{GMMs}) 
to describe their statistical distribution. Then, they trained \Acronym{HMMs} to account for the temporal evolution of the \Acronym{GMMs} using a discriminative algorithm that exploited knowledge about the categories of training signals
. Eronen and co-authors  \cite{Eronen2006Au} further developed on this work by considering a larger group of features, and by adding a feature transform step 
to the classification algorithm, obtaining an overall $58\%$ accuracy in the classification of $18$ different acoustic scenes.

In the algorithms mentioned so far, each signal belonging to a training set of recordings is generally divided into frames of fixed duration, and a transform is applied to each frame to obtain a sequence of \emph{feature vectors}. The feature vectors derived from each acoustic scene are then employed to train a \emph{statistical model} that summarises the properties of a whole soundscape, or of multiple soundscapes belonging to the same category. Finally, a \emph{decision criterion} is defined to assign unlabelled recordings to the category that best matches the distribution of their features. A more formal definition of an \Acronym{ASC} framework will be presented in Section \ref{sec:fra}, and the details of a signal processing challenge we have organised to benchmark \Acronym{ASC} methods will be presented in Section \ref{sec:peda}. Here we complete the historical overview of computational \Acronym{ASC} and emphasise their main contributions in light of the components identified above.

\subsection{Features}\label{sse:ftm}
Several categories of audio features have been employed in \Acronym{ASC} systems. Here we present a list of them, providing their rationale in the context of audio analysis for classification.
\subsubsection{Low-level time-based and frequency-based audio descriptors} several \Acronym{ASC} systems \cite[GSR]{DCASE}\textcolor{red}{\footnote{Here and throughout the paper the notation [1, XXX] is used to cite the extended abstracts submitted for the \Acronym{DCASE} challenge described in Section \ref{sec:peda}. The code XXX (e.g., ``GSR'')  corresponds to a particular submission to the challenge (see Table \ref{tab:dcase}).}}\cite{Eronen2006Au,Malkin2005CL} employ features that can be easily computed from either the signal in the time domain or its Fourier transform. These include (among others) the \emph{zero crossing rate} which measures the average rate of sign changes within a signal, and is related to the main frequency of a monophonic sound; the \emph{spectral centroid}, which measures the centre of mass of the spectrum and it is related to the perception of \emph{brightness} \cite{Grey1978Pe}; and the \emph{spectral roll-off} that identifies a frequency above which the magnitude of the spectrum falls below a set threshold.
\subsubsection{Frequency-band energy features (energy/frequency)} this class of features used by various \Acronym{ASC} systems \cite[NR CHR GSR]{DCASE}\cite{Eronen2006Au} is computed by integrating the magnitude spectrum or the power spectrum over specified frequency bands. The resulting coefficients measure the amount of energy present within different sub-bands, and can also be expressed as a ratio between the sub-band energy and the total energy to encode the most prominent frequency regions in the signal.
\subsubsection{Auditory filter banks} A further development of energy/frequency features consists in analysing audio frames through filter banks that mimic the response of the human auditory system. Sawhney and Maes \cite{Sawhney1997Si} used Gammatone filters for this purpose, Clarkson \emph{et al.} \cite{Clarkson1998Au} instead computed Mel-scaled filter bank coefficients (\Acronym{MFCs}), whereas Patil and Elahili \cite[PE]{DCASE} employed a so-called auditory spectrogram.
\subsubsection{Cepstral features} \Acronym{MFCCs} are an example of cepstral features and are perhaps the most popular features used in \Acronym{ASC}. They are obtained by computing the discrete cosine transform (\Acronym{DCT}) of the logarithm of \Acronym{MFCs}. The name \emph{cepstral} is an anagram of \emph{spectral}, and indicates that this class of features is computed by applying a Fourier-related transform to the spectrum of a signal. Cepstral features capture the spectral envelope of a sound, and thus summarise their coarse spectral content.
\subsubsection{Spatial features} If the soundscape has been recorded using multiple microphones, features can be extracted from the different channels to capture properties of the acoustic scene. In the case of a stereo recording, popular features include the \emph{inter-aural time difference} (\Acronym{ITD}) that measures the relative delay occurring between the left and right channels when recording a sound source; and the \emph{inter-aural level difference} (\Acronym{ILD}) measuring the amplitude variation between channels. Both \Acronym{ITD} and \Acronym{ILD} are linked to the position of a sound source in the stereo field. Nogueira \emph{et al.} \cite[NR]{DCASE} included spatial features in their \Acronym{ASC} system.
\subsubsection{Voicing features} Whenever the signal is thought to contain harmonic components, a fundamental frequency $f_0$ or a set of fundamental frequencies can be estimated, and groups of features can be defined to measure properties of these estimates. In the case of \Acronym{ASC}, harmonic components might correspond to specific events occurring within the audio scene, and their identification can help discriminate between different scenes. Geiger \emph{et al.} \cite[GSR]{DCASE} employed voicing features related to the fundamental frequency of each frame in their system. The method proposed by Krijnders and Holt \cite[KH]{DCASE} is based on extracting tone-fit features, a sequence of voicing features derived from a perceptually motivated representation of the audio signals. Firstly, a so-called cochleogram  is computed to provide a time-frequency representation of the acoustic scenes that is inspired by the properties of the human cochlea. Then, the \emph{tonalness} of each time-frequency region is evaluated to identify tonal events in the acoustic scenes, resulting in tone-fit feature vectors.
\subsubsection{Linear predictive coefficients (\Acronym{LPCs})} this class of features have been employed in the analysis of speech signals that are modelled as autoregressive processes. In an autoregressive model, samples of a signal $\sig$ at a given time instant $\iTim$ are expressed as linear combinations of samples at \textcolor{red}{$\nLags$} previous time instants:
\begin{equation}
	\sig(\iTim) = \sum_{\iLag=1}^{\nLags}\lpcCoe_{\iLag}\sig(\iTim-\iLag) + \epsilon(\iTim)
\end{equation}
where the combination coefficients $\curlyb{\lpcCoe_{\iLag}}_{\iLag=1}^{\nLags}$ determine the model parameters and $\epsilon$ is a residual term. There is a mapping between the value of \Acronym{LPCs} and the spectral envelope of the modelled signal \cite{Rabiner1993Fu}, therefore $\lpcCoe_{\iLag}$ encode information regarding the general spectral characteristics of a sound. Eronen \emph{et al.} \cite{Eronen2006Au} employed \Acronym{LPC} features in their proposed method.
\subsubsection{Parametric approximation features}autoregressive models are a special case of approximation models where a signal $\sig$ is expressed as a linear combination of \textcolor{red}{$\nAtos$} basis functions from the set $\curlyb{\ato_{\iAto}}_{\iAto=1}^{\nAtos}$
\begin{equation}\label{eq:lm}
	\sig(\iTim) = \sum_{\iAto=1}^{\nAtos}\lpcCoe_{\iAto}\ato_{\iAto}(\iTim) + \epsilon(\iTim).
\end{equation}
Whenever the basis functions $\ato_{\iAto}$ are parametrized by a set of parameters $\gamma_{\iAto}$, features can be defined according to the functions that contribute to the approximation of the signal. For example, Chu \emph{et al.} \cite{Chu2009En} decompose audio scenes using the Gabor transform, that is a  representation where each basis function is parametrized by its frequency $f$, its time scale $u$, its time shift $\tau$ and its frequency phase $\theta$; so that $\gamma_{\iAto}=\curlyb{f_{\iAto},u_{\iAto},\tau_{\iAto},\theta_{\iAto}}$. The set of indexes identifying non-zero coefficients $\optimal{\iAto} = \curlyb{\iAto : \lpcCoe_{\iAto}\neq 0}$ corresponds to a set of active parameters $\gamma_{\optimal{\iAto}}$ contributing to the approximation of the signal, and encode events in an audio scene that occur at specific time-frequency locations. Patil and Elahili \cite[PE]{DCASE} also extract parametric features derived from the 2-dimensional convolution between the auditory spectrogram and 2D Gabor filters.

\subsubsection{Unsupervised learning features} The model \eqref{eq:lm} assumes that a set of basis functions is defined \emph{a priori} to analyse a signal. Alternatively, bases can be learned from the data or from other features already extracted in an unsupervised way. Nam \emph{et al.} \cite[NHL]{DCASE} employed a sparse restricted Boltzman machine (\Acronym{SRBM}) to adaptively learn features from the \Acronym{MFCCs} of the training data. A \Acronym{SRBM} is a neural network that has been shown to learn basis functions from input images which resemble the properties of representations built by the visual receptors in the human brain. In the context of \Acronym{ASC}, a \Acronym{SRBM} adaptively encodes basic properties of the spectrum of the training signals and returns a sequence of features learned from the \Acronym{MFCCs}, along with an activation function that is used to determine time segments containing significant acoustic events.
\subsubsection{Matrix factorisation methods} The goal of matrix factorisation for audio applications is to describe the spectrogram of an acoustic signal as a linear combination of elementary functions that capture typical or salient spectral elements, and are therefore a class of unsupervised learning features. The main intuition that justifies using matrix factorisation for classification is that the signature of events that are important in the recognition of an acoustic scene should be encoded in the elementary functions, leading to discriminative learning. Cauchi \cite{Cauchi2011No} employed non-negative matrix factorisation (\Acronym{NMF}) and Benetos \emph{et al.} \cite{Benetos2012Ch} used probabilistic latent component analysis in their proposed algorithms. Note that a matrix factorisation also outputs a set of activation functions which encode the contribution of elementary functions in time, hence modelling the properties of a whole soundscape. Therefore, this class of techniques can be considered to jointly estimate local and global parameters.

\subsubsection{Image processing features} Rakotomamonjy and Gasso \cite[RG]{DCASE} designed an algorithm for \Acronym{ASC} whose feature extraction function comprises the following operations. Firstly, the audio signals corresponding to each training scene are processed using a constant-Q transform, which returns frequency representations with logarithmically-spaced frequency bands. Then, $512\times 512$-pixel grayscale images are obtained from the constant-Q representations by interpolating neighbouring time-frequency bins. Finally, features are extracted from the images by computing the matrix of local gradient histograms. This is obtained by dividing the images into local patches, by defining a set of spatial orientation directions, and by counting the occurrence of edges exhibiting each orientation. Note that in this case the vectors of features are not independently extracted from frames, but from time-frequency tiles of the constant-Q transform.

\subsubsection{Event detection and acoustic unit descriptors} Heittola \emph{et al.} \cite{Heittola2010Au} proposed a system for \Acronym{ASC} that classifies soundscapes based on a histogram of events detected in a signal. During the training phase, the occurrence of manually annotated events (such as 'car horn', 'applause' or 'basketball')  is used to derive models for each scene category. In the test phase, \Acronym{HMMs} are employed to identify events within an unlabelled recording, and to define a histogram that is compared to the ones derived from the training data. This system represents an alternative to the common framework that includes features, statistical learning and a decision criterion, in that it essentially performs event detection and \Acronym{ASC} at the same time. However, for the purpose of this tutorial, the acoustic events can be thought as high-level features whose statistical properties are described by histograms. 

A similar strategy is employed by Chauduri \emph{et al.} \cite{Chaudhuri2011Un} to learn acoustic unit descriptors (\Acronym{AUDs}) and classify YouTube multimedia data. \Acronym{AUDs} are modelled using \Acronym{HMMs}, and used to transcribe an audio recording into a sequence of events. The transcriptions are assumed to be generated by N-gram language models whose parameters are trained on different soundscapes categories. The transcriptions of unlabelled recordings during the test phase are thus classified following a maximum likelihood criterion.

\textcolor{red}{\subsection{Feature processing}
The features described so far can be further processed to derive new quantities that are used either in place or as an addition to the original features.}

\subsubsection{Feature transforms} This class of methods is used to enhance the discriminative capability of features by processing them through linear or non-linear transforms. Principal component analysis (\Acronym{PCA}) is perhaps the most commonly cited example of feature transforms. It learns a set of orthonormal basis that minimise the Euclidean error that results from projecting the features onto subspaces spanned by subsets of the basis set (the principal components), and hence identifies the directions of maximum variance in the dataset. Because of this property, \Acronym{PCA} (and the more general independent component analysis (\Acronym{ICA})) have been employed as a dimensionality reduction technique to project high-dimensional features onto lower dimensional subspaces while retaining the maximum possible amount of variance \cite[PE]{DCASE}\cite{Eronen2006Au,Malkin2005CL}. Nogueira \emph{et al.} \cite[NR]{DCASE}, on the other hand, evaluate a Fisher score to measure how features belonging to the same class are clustered near each other and far apart from features belonging to different classes. A high Fisher score implies that features extracted from different classes are likely to be separable, and it is used to select optimal subsets of features.

\subsubsection{Time derivatives} For all the quantities computed on local frames, discrete time derivatives between consecutive frames can be included as additional features that identify the time evolution of the properties of an audio scene.
	
Once features are extracted from audio frames, the next stage of an \Acronym{ASC} system generally consists of learning statistical models of the distribution of the features.

\subsection{Statistical models}\label{sec:sm}
Statistical models are parametric mathematical models used to summarise the properties of individual audio scenes or whole soundscape categories from the feature vectors. They can be divided into \emph{generative} or \emph{discriminative} methods.

When working with generative models, feature vectors are interpreted as being generated from one of a set of underlying statistical distributions. During the training stage, the parameters of the distributions are optimised based on the statistics of the training data. In the test phase, a decision criterion is defined to determine the most likely model that generated a particular observed example. A simple implementation of this principle is to compute basic statistical properties of the distribution of feature vectors belonging to different categories (such as their mean values), hence obtaining one class \emph{centroid} for each category. The same statistic can be computed for each unlabelled sample  that is assumed to be generated according to the distribution with the closest centroid, and is assigned to the corresponding category. 



When using a discriminative classifier, on the other hand, features derived from an unlabelled sample are not interpreted as being generated by a class-specific distribution, but are assumed to occupy a class-specific region in the feature space. One of the most popular discriminative classifiers for \Acronym{ASC} is the support vector machine (\Acronym{SVM}). The model output from an \Acronym{SVM} determines a set of hyperplanes that optimally separate features associated to different classes in the training set (according to a maximum-margin criterion). An \Acronym{SVM} can only discriminate between two classes. However, when the classification problem includes more than two categories (as is the case of the \Acronym{ASC} task presented in this paper), multiple \Acronym{SVMs} can be combined to determine a decision criterion that allows to discriminate between $\nUniCat$ classes. In the \emph{one versus all} approach, $\nUniCat$ \Acronym{SVMs} are trained to discriminate between data belonging to one class and data from the remaining $\nUniCat-1$ classes.  Instead, in the \emph{one versus one} approach $\nUniCat(\nUniCat-1)/2$ \Acronym{SVMs} are trained to classify between all the possible class combinations. In both cases, the decision criterion estimates the \textcolor{red}{class} from an unlabelled sample by evaluating the distance between the data and the separating hyperplanes learned by the \Acronym{SVMs}.

Discriminative models can be combined with generative ones. For example, one might use the parameters of generative models learned from training data to define a feature space, and then employ an \Acronym{SVM} to learn separating hyperplanes. In other words, discriminative classifiers can be used to derive classification criteria from either the feature vectors or from the parameters of their statistical models. In the former case, the overall classification of an acoustic scene must be decided from the classification of individual data frames using, for example, a majority vote.

Different statistical models have been used for computational \Acronym{ASC}, and the following list highlights their categories.
\subsubsection{Descriptive statistics} Several techniques for \Acronym{ASC} \cite[KH GSR RNH]{DCASE} employ descriptive statistics. This class of methods is used to quantify various aspects of statistical distributions, including moments (such as mean, variance, skewness and kurtosis of a distribution), quantiles and percentiles. \subsubsection{Gaussian mixture models (\Acronym{GMMs})} Other methods for \Acronym{ASC} \cite{Chu2006WH,Aucouturier2007Th} employ \Acronym{GMMs}, that are generative methods where feature vectors are interpreted as being generated by a multi-modal distribution expressed as a sum of Gaussian distributions. \Acronym{GMMs} will be further detailed in Section \ref{sse:a-b} where we will present a baseline \Acronym{ASC} system used for benchmark.
\subsubsection{Hidden Markov Models (\Acronym{HMMs})} This class of models are used in several \Acronym{ASC} systems \cite{Clarkson1998Au,Eronen2006Au} to account for the temporal unfolding of events within complex soundscapes. Suppose, for example, that an acoustic scene recorded in an underground train includes an alert sound preceding the sound of the doors closing and the noise of the electric motor moving the carriage to the next station. Features extracted from these three distinct sounds could be modelled using Gaussian densities with different parameters, and the order in which the events normally occur would be encoded in an \Acronym{HMM} transition matrix. This contains the transition probability between different states at successive times, that is the probability of each sound occurring after each other. A transition matrix that correctly models the unfolding of events in an underground train would contain large diagonal elements indicating the probability of sounds persisting in time, significant probabilities connecting events that occur after each other (the sound of motors occurring after the sound of doors occurring after the alert sound), and negligible probabilities connecting sounds that occur in the wrong order (for example the doors closing before the alert sound).
\subsubsection{Recurrence quantification analysis (\Acronym{RQA})} Roma \emph{et al.}   \cite[RNH]{DCASE} employ \Acronym{RQA} to model the temporal unfolding of acoustic events. This technique is used to learn a set of parameters that have been developed to study dynamical systems in the context of chaos theory, and are derived from so-called recurrence plots which capture periodicities in a time series. In the context of \Acronym{ASC}, the \Acronym{RQA} parameters include: \emph{recurrence} measuring the degree of self-similarity  of features within an audio scene; \emph{determinism} which is correlated to sounds periodicities and \emph{laminarity} that captures sounds containing stationary segments.
The outputs of the statistical learning function are a set of parameters that model each acoustic scene in the training set. This collection of parameters is then fed to an \Acronym{SVM} to define decision boundaries between classes that are used to classify unlabelled signals.

\subsubsection{i-vector} The system proposed by Elizalde \emph{et al.} \cite[ELF]{DCASE} is based on the computation of the \emph{i-vector} \cite{Dehak2009Su}. This is a technique originally developed in the speech processing community to address a speaker verification problem, and it is based on modelling a sequence of features using \Acronym{GMMs}. In the context of \Acronym{ASC}, the i-vector is specifically derived as a function of the parameters of the \Acronym{GMMs} learned from \Acronym{MFCCs}. It leads to a low-dimensional representation summarising the properties of an acoustic scene, and is input to a generative probabilistic linear discriminant analysis (\Acronym{pLDA}) \cite{Ioffe2006Pr}.  

\subsection{Decision criteria}\label{sec:dc}
Decision criteria are functions used to determine the category of an unlabelled sample from its feature vectors and from the statistical model learned from the set of training samples. Decisions criteria are generally dependent on the type of statistical learning methods used, and the following list details how different models are associated to the respective criteria.
\subsubsection{One-vs-one and one-vs-all} this pair of decision criteria are associated to the output of a multi-class \Acronym{SVM}, and are used to map the position of a features vector to a class, as already described in Section \ref{sec:sm}.
\subsubsection{Majority vote} This criterion is used whenever a global classification must be estimated from decisions about single audio frames. Usually, an audio scene is classified according to the most common category assigned to its frames. Alternatively, a weighted majority vote can be employed to vary the importance of different frames. Patil and Elahili \cite[PE]{DCASE}, for example, assign larger weights to audio frames \textcolor{red}{containing more energy}.
\subsubsection{Nearest neighbour} According to this criterion, a feature vector is assigned to the class associated to the closest vector from the training set (according to a metric, often the Euclidean distance). A generalisation of nearest neighbour is the \emph{k-nearest neighbour} criterion, whereby the $k$ closest vectors are considered and a category is determined according to the most common classification.
\subsubsection{Maximum likelihood} This criterion is associated with generative models, whereby feature vectors are assigned to the category whose model is most likely to have generated the observed data according to a likelihood probability. 
\subsubsection{Maximum a posteriori}
An alternative to maximum likelihood classification is the \emph{maximum a posteriori (\Acronym{MAP})} criterion that includes information regarding the  marginal likelihood of any given class. For instance, suppose that a \Acronym{GPS} system in a mobile device indicates that in the current geographic area some environments are more likely to be encountered than others. This information could be included in an \Acronym{ASC} algorithm through a \Acronym{MAP} criterion.

\subsection{Meta-algorithms} 
\textcolor{red}{In the context of supervised classification, meta-algorithms are machine learning techniques designed to reduce the classification error by running multiple instances of a classifier in parallel, each of which uses different parameters or different training data. The results of each classifier are then combined into a global decision.}


\subsubsection{Decision trees and tree-bagger} A decision tree is a set of rules derived from the analysis of features extracted from training signals. It is an alternative to generative and discriminative models because it instead optimises a set of \emph{if/else} conditions about the values of features that leads to a classification output. Li \emph{et al.} \cite[LTT]{DCASE} employed a tree-bagger classifier, that is a set of multiple decision trees. A tree-bagger is an example of a classification meta-algorithm that computes multiple so called \emph{weak learners} (classifiers whose accuracy is only assumed to be better than chance) from randomly-sampled copies of the training data following a process called bootstrapping. In the method proposed by Lee \emph{et al.} the ensemble of weak learners are then combined to determine a category for each frame, and in the test phase an overall category is assigned to each acoustic scene based on a majority vote.

\subsubsection{Normalized compression dissimilarity and random forest} Olivetti \cite[OE]{DCASE} adopts a system for \Acronym{ASC} that departs from techniques described throughout this paper in favour of a method based on audio compression and random forest. Motivated by the theory of Kolmogorov complexity which measures the shortest binary program that outputs a signal, and that is approximated using compression algorithms, he defines a normalised compression distance between two audio scenes. This is a function of the size in bits of the files obtained by compressing the acoustic scenes using any suitable audio coder. From the set of pairwise distances, a classification is obtained using a random forest, that is a meta-algorithm based on decision trees.

\subsubsection{Majoriy vote and boosting}
The components of a classification algorithm can be themselves thought as parameters subject to optimisation. Thus, a further class of meta-algorithms deals with selecting from or combining multiple classifiers to improve the classification accuracy. Perhaps the simplest implementation of this general idea is to run several classification algorithms in parallel on each test sample and determine the optimal category by majority vote, an approach that will be also used in Section \ref{sec:sp} of this article. Other more sophisticated methods include \emph{boosting} techniques \cite{Schapire2003Th} where the overall classification criterion is a function of linear combinations involving a set of weak learners.

\section{A general framework for \Acronym{ASC}}\label{sec:fra}
Now that we have seen the range of machine learning and signal processing techniques used in the context of \Acronym{ASC}, let us define a framework that allows us to distill a few key operators and components.
Computational algorithms for \Acronym{ASC} are designed to solve a supervised classification problem where a set of $\nSig$ training recordings $\curlyb{\sig_{\iSig}}_{\iSig=1}^{\nSig}$ is provided and associated with corresponding labels $\curlyb{\cat_{\iCat}}_{\iCat=1}^{\nCat}$ that indicate the category to which each soundscape belongs. Let $\curlyb{\uniCat_{\iUniCat}}_{\iUniCat=1}^{\nUniCat}$ be a set of labels indicating the members of a universe of $\nUniCat$ possible categories. Each label $\cat_{\iSig}$ can assume one of the values in this set, and we define a set $\idxSet_{\iUniCat}=\curlyb{\iSig : \cat_{\iSig}=\uniCat_{\iUniCat}}$ that identifies the signals belonging to the $\iUniCat$-th class. The system learns statistical models from the different classes during an off-line training phase, and uses them to classify unlabelled recordings $\sig_{\new}$ in the test phase.

Firstly, each of the training signals is divided into short frames. Let $\nFraDim$ be the length of each frame, $\sig_{\iFra,\iSig}\in\real^{\nFraDim}$ indicates the $\iFra$-th frame of the $\iSig$-th signal.  Typically, $\nFraDim$ is chosen so that the frames duration is about $50$ms depending on the signal's sampling rate.

Frames in the time domain are not directly employed for classification, but are rather used to extract a sequence of features through a transform $\tra : \tra(\sig_{\iFra,\iSig}) = \fea_{\iFra,\iSig}$, where $\fea_{\iFra,\iSig}\in\real^{\nFeaDim}$ indicates a vector of features of dimension $\nFeaDim$. Often, $\nFeaDim\ll\nFraDim$ meaning that $\tra$ causes a dimensionality reduction. This is aimed at obtaining a coarser representation of the training data where members of the same class  result in similar features (yielding \emph{generalisation}), and members of different classes can be distinguished from each other (allowing \emph{discrimination}). Some systems further manipulate the features using feature transforms, such as in the method proposed by Eronen \emph{et al.} \cite{Eronen2006Au}. For clarity of notation, we will omit this additional feature processing step from the description of the \Acronym{ASC} framework, considering any manipulation of the features to be included in the operator $\tra$.

\begin{figure}[!tb]
\begin{center}
\includegraphics[width=\textwidth]{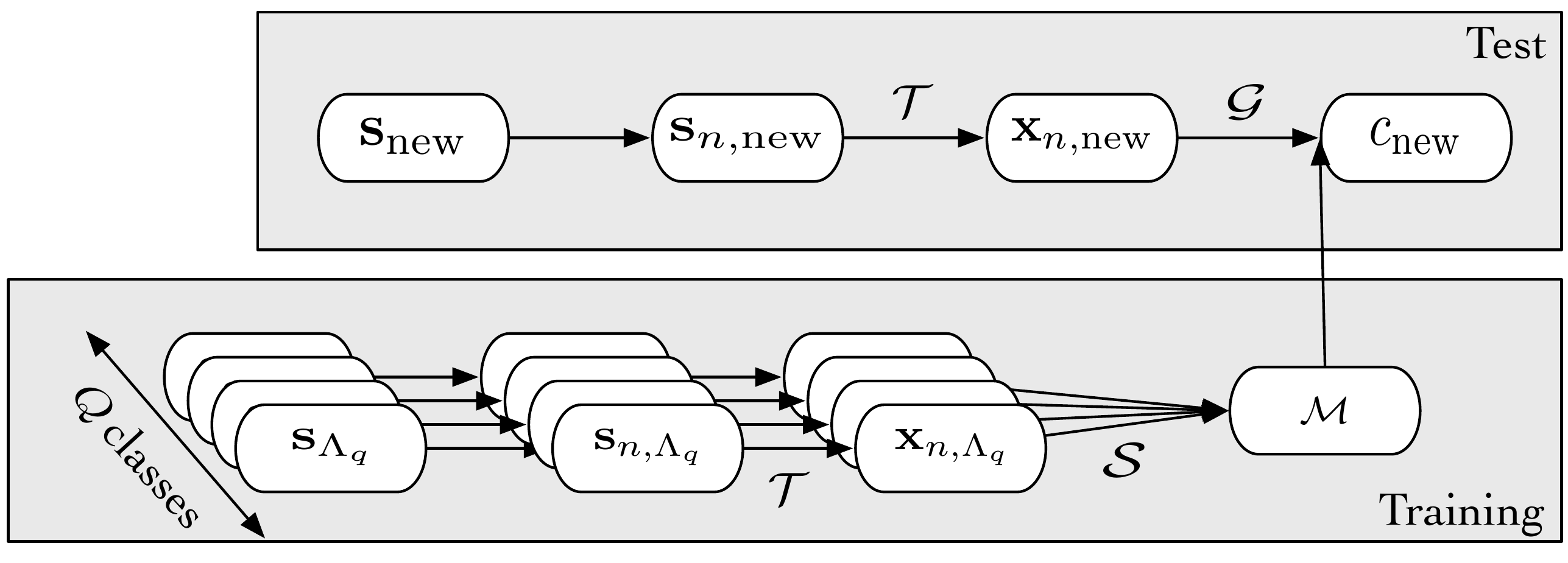}
\caption{\label{fig:fra}Supervised classification framework for acoustic scene classification.}
\end{center}
\end{figure}

Individual features obtained from time-localised frames cannot summarise the properties of soundscapes that are constituted by a number of different events occurring at different times. For this reason, sequences of features extracted from signals belonging to a given category are used to learn statistical models of that category, abstracting the classes from their empirical realisations.  Let $\fea_{\iFra,\idxSet_{\iUniCat}}$ indicate the features extracted from the signals belonging to the $\iUniCat$-th category. The function $\lea : \lea\roundb{\curlyb{\fea_{\iFra,\idxSet_{\iUniCat}}}} = \model$ learns the parameters of a statistical model $\model$ that describes the global properties of the training data. Note that this formulation of the statistical learning stage (also illustrated in Figure \ref{fig:fra}) can describe a discriminative function that requires features from the whole training set to compute separation boundaries between classes. In the case of generative learning, the output of the function $\lea$ can be separated into $\nUniCat$ independent models $\curlyb{\model_{\iUniCat}}$ containing parameters for each category, or into $\nSig$ independent models $\curlyb{\model_{\iSig}}$ corresponding to each training signal.  

Once the training phase has been completed, and a model $\model$ has been learned, the transform $\tra$ is applied in the test phase to a new unlabelled recording $\sig_{\new}$, leading to a sequence  of features $\fea_{\new}$. A function $\dec : \dec(\fea_{\new},\model) = \cat_{\new}$ is then employed to classify the signal, returning a label in the set $\curlyb{\uniCat_{\iUniCat}}_{\iUniCat=1}^{\nUniCat}$.

Most of the algorithms mentioned in Section \ref{sec:back} follow the framework depicted in Figure \ref{fig:fra}, and only differ in their choice of the functions $\tra$, $\lea$ and $\dec$. Some follow a seemingly different strategy, but can still be analysed in light of this framework: for example, matrix factorisations algorithms like the one proposed by Benetos \emph{et al.} \cite{Benetos2012Ch} can be interpreted as combining features extraction and statistical modelling through the unsupervised learning of spectral templates and an activation matrix, as already discussed in Section \ref{sse:ftm}.

A special case of \Acronym{ASC} framework is the so-called \emph{bag-of-frames} approach \cite{Aucouturier2007Th}, named in an analogy with the \emph{bag-of-words} technique for text classification whereby documents are described by the distribution of their word occurrences. Bag-of-frames techniques follow the general structure shown in Figure \ref{fig:fra}, but ignore the ordering of the sequence of features when learning statistical models.

\section{Challenge on Detection and classification of acoustic scenes and events}\label{sec:peda}
Despite a rich literature on systems for \Acronym{ASC}, the research community has so far lacked a coordinated effort to evaluate and benchmark algorithms that tackle this problem. The challenge on detection and classification of acoustic scenes and events (\Acronym{DCASE}) has been organised in partnership with the \Acronym{IEEE} Audio and Acoustic Signal Processing (\Acronym{AASP}) Technical Committee in order to test and compare algorithms for \Acronym{ASC} and for event detection and classification. This initiative is in line with a wider trend in the signal processing community aimed at promoting reproducible research \cite{vandewalle2009}. Similar challenges have been organised in the areas of music information retrieval \cite{MIREX}, speech recognition \cite{Barker:2012a} and source separation \cite{SiSEC}.

\subsection{The \Acronym{DCASE} dataset}
Existing algorithms for \Acronym{ASC} have been generally tested on datasets that are not publicly available \cite{Sawhney1997Si,Eronen2006Au}, making it difficult if not impossible to produce sustainable and reproducible experiments built on previous research. Creative-commons licensed sounds can be accessed for research purposes on freesound.org\footnote{\href{http://freesound.org}{http://freesound.org}}, a collaborative database that includes environmental sounds along with music, speech and audio effects. However, the different recording conditions and varying quality of the data present in this repository would require a substantial curating effort to identify a set of signals suited for a rigorous and fair evaluation of \Acronym{ASC} systems. On the other hand, the adoption of commercially available databases such as the Series 6000 General Sound Effects Library\footnote{\href{http://www.sound-ideas.com/sound-effects/series-6000-sound-effects-library.html}{http://www.sound-ideas.com/sound-effects/series-6000-sound-effects-library.html}} would constitute a barrier to research reproducibility due to their purchase cost.

The \Acronym{DCASE} challenge dataset \cite{Giannoulis2013A-} was especially created to provide researchers with a standardised set of recordings produced in 10 different urban environments. The soundscapes have been recorded in the London area and include: `bus', `busy-street', `office', `openairmarket', `park', `quiet-street', `restaurant', `supermarket', `tube' (underground railway) and `tubestation'. Two disjoint datasets were constructed from the same group of recordings each containing ten $30$s long clips for each scene, totalling $100$ recordings. Of these two datasets, one is publicly available and can be used by researchers to train and test their \Acronym{ASC} algorithms; the other has been held-back and has been used to evaluate the methods submitted for the challenge.

\subsection{List of submissions}
A total of $11$ algorithms were proposed for the \Acronym{DCASE} challenge on \Acronym{ASC} from research institutions worldwide. The respective authors submitted accompanying extended abstracts describing their techniques which can be accessed from the \Acronym{DCASE} website \footnote{\href{http://c4dm.eecs.qmul.ac.uk/sceneseventschallenge/}{http://c4dm.eecs.qmul.ac.uk/sceneseventschallenge/}}. The following table lists the authors and titles of the contributions, and defines acronyms that are used throughout the paper to refer to the algorithms.

In addition to the methods submitted for the challenge, we designed a benchmark baseline system that employes \Acronym{MFCCs}, \Acronym{GMMs} and a maximum likelihood criterion. \textcolor{red}{We have chosen to use these components because they represent standard practices in audio analysis which are not specifically tailored to the \Acronym{ASC} problem, and therefore provide an interesting comparison with more sophisticated techniques.}
\def \textWidthPerc{0.3}
\begin{longtable}{|p{.09\textwidth}|p{.28\textwidth}|p{.55\textwidth}|} 
\hline
\textbf{Acronym} & \textbf{Authors} & \textbf{Title} \\
\hline
\Acronym{RNH} & G.~Roma, W. Nogueira and P.~Herrera & Recurrence quantification analysis features for auditory scene classification \\
\hline
\Acronym{RG} & A. Rakotomamonjy and G.~Gasso & Histogram of gradients of time-frequency representations for audio scene classification\\
\hline
\Acronym{GSR} & J. T. Geiger, B. Schuller and G. Rigoll & Recognising acoustic scenes with large-scale audio feature extraction and \Acronym{SVM} \\
\hline
\Acronym{CHR} & M. Chum, A. Habshush, A.~Rahman and C. Sang & \Acronym{IEEE AASP} Scene classification challenge using hidden Markov models and frame based classification \\
\hline
\Acronym{NHL} & J. Nam, Z. Hyung and K. Lee & Acoustic scene classification using sparse feature learning and selective max-pooling by event detection \\
\hline
\Acronym{NR} & W. Nogueira, G. Roma, and P. Herrera & Sound scene identification based on \Acronym{MFCC}, binaural features and a support vector machine classifier\\
\hline
\Acronym{PE} & K. Patil and M. Elhilali & Multiresolution auditory representations for scene classification\\
\hline
\Acronym{KH} & J. Krijnders and G. A. T. Holt & A tone-fit feature representation for scene classification\\
\hline
\Acronym{ELF} & B. Elizalde H. Lei, G. Friedland and N. Peters & An I-vector based approach for audio scene detection\\
\hline
\Acronym{LTT}\footnote{The original \Acronym{LTT} submission achieved low accuracy due to a bug in a Matlab toolbox - here we are presenting the results obtained with the correct implementation.} & David Li, Jason Tam, and Derek Toub & Auditory scene classification using machine learning techniques\\
\hline
\Acronym{OE} & E. Olivetti & The wonders of the normalized compression dissimilarity representation\\
\hline
\caption{\label{tab:dcase} List of algorithms submitted for the \Acronym{DCASE} challenge on \Acronym{ASC}.} 
\end{longtable}

\section{Summary table of algorithms for \Acronym{ASC}}
Having described the \Acronym{ASC} framework in Section \ref{sec:fra} and  the methods submitted for the \Acronym{DCASE} challenge throughout Section \ref{sec:back} and in Section \ref{sec:peda}, we now present a table that summarises the various approaches.

\def \textWidthPerc{0.35}
\begin{landscape}
\begin{longtable}{|p{.15\textwidth}|p{\textWidthPerc\textwidth}|p{\textWidthPerc\textwidth}|p{\textWidthPerc\textwidth}|} 
\hline
\textbf{Method} & \textcolor{red}{Features} & \textcolor{red}{Statistical model} & \textcolor{red}{Decision criterion} \\
\hline
Sawhney and Maes \cite{Sawhney1997Si} & Filter bank & None & Nearest neighbour $\rightarrow$ majority vote \\
\hline
Clarkson \emph{et al.} \cite{Clarkson1998Au} & \Acronym{MFCs} & \Acronym{HMM} & Maximum likelihood \\
\hline
Eronen \emph{et al.} \cite{Eronen2006Au} & \Acronym{MFCCs}, low-level descriptors, energy/frequency, \Acronym{LPCs} $\rightarrow$ \Acronym{ICA}, \Acronym{PCA} & Discriminative \Acronym{HMM} & Maximum likelihood\\
\hline
Aucouturier \cite{Aucouturier2007Th} & \Acronym{MFCCs} & \Acronym{GMMs} & Nearest neighbour\\
\hline
Chu \emph{et al.} \cite{Chu2009En} & \Acronym{MFCCs}, parametric (Gabor) & \Acronym{GMMs} & Maximum likelihood\\
\hline
Malkin and Waibel \cite{Malkin2005CL} & \Acronym{MFCCs}, low-level descriptors $\rightarrow$ \Acronym{PCA} & Linear auto-encoder networks & Maximum likelihood \\
\hline
Cauchi \cite{Cauchi2011No} & \Acronym{NMF} &  & Maximum likelihood \\
\hline
Benetos \cite{Benetos2012Ch} & \Acronym{PLCA} &  & Maximum likelihood\\
\hline
Heittola \emph{et al.} \cite{Heittola2010Au}& Acoustic events & Histogram & Maximum likelihood \\
\hline
Chaudhuri \emph{et al.} \cite{Chaudhuri2011Un} & Acoustic unit descriptors & N-gram language models & Maximum likelihood \\
\hline
\multicolumn{4}{c}{\textbf{\Acronym{DCASE} Submissions}} \\
\hline
Baseline & \Acronym{MFCCs} & \Acronym{GMMs} & Maximum likelihood \\
\hline
\Acronym{RNH}
& \Acronym{MFCCs} & \Acronym{RQA}, moments $\rightarrow$ \Acronym{SVM} & - \\
\hline
\Acronym{RG}
& Local gradient histograms (learned on time-frequency patches) & Aggregation $\rightarrow$ \Acronym{SVM} & One versus one \\
\hline
\textbf{Method} & \textcolor{red}{Features} & \textcolor{red}{Statistical model} & \textcolor{red}{Decision criterion} \\
\hline
\Acronym{GSR}
& \Acronym{MFCCs}, energy/frequency, voicing & Moments, percentiles, linear regression coeff. $\rightarrow$ \Acronym{SVM} & Majority vote \\
\hline
\Acronym{CHR}
& Energy/frequency & \Acronym{SVM} & One versus all, majority vote \\
\hline
\Acronym{NHL}
& Learned (\Acronym{MFCCs} $\rightarrow$  \Acronym{SRBM}) & Selective max pooling $\rightarrow$ \Acronym{SVM} & One versus all\\
\hline
\Acronym{NR}
& \Acronym{MFCCs}, energy/frequency, spatial $\rightarrow$ Fisher feature selection & \Acronym{SVM} & Majority vote\\
\hline
\Acronym{PE}
& filter bank $\rightarrow$ parametric (Gabor)  $\rightarrow$ \Acronym{PCA} & \Acronym{SVM} & One versus one, weighted majority vote\\
\hline
\Acronym{KH}
& Voicing & Moments, percentiles $\rightarrow$ \Acronym{SVM} & -\\
\hline
\Acronym{ELF}
& \Acronym{MFCCs} & i-vector  $\rightarrow$ \Acronym{pLDA} & Maximum likelihood\\
\hline
\Acronym{LTT}
& \Acronym{MFCCs} & Ensemble of classification trees &  majority vote $\rightarrow$ treebagger \\
\hline
\Acronym{OE}
& Size of compressed audio & Compression distance -> ensemble of classification trees & - $\rightarrow$ random forest\\
\hline
\caption{\label{tab:asc}Summary and categorisation of computational methods for \Acronym{ASC}. The acronyms after the author(s) name(s) in the method column are defined in Table \ref{tab:dcase}. Arrows indicate sequential processing, for example when statistical parameters learned from features are fed to an \Acronym{SVM} to obtain separating hyperplanes. In some cases the decision criterion of \Acronym{SVMs} (one versus all, one versus one, or alternative) is not specified in the reference. However, it is always specified when the discriminative learning is performed on frames and an overall classification is determined by a majority vote or a weighted majority vote. Note that for each work cited only the method leading to best classification results have been considered.} 
\label{tab:myfirstlongtable}
\end{longtable}
\end{landscape}

\section{Evaluation of algorithms for \Acronym{ASC}}\label{sec:sp}
\subsection{Experimental design}
A system designed for \Acronym{ASC} comprises training and test phases. Researchers who participated to the \Acronym{DCASE} challenge were provided with a public dataset that includes ground truth labels indicating the environment in which sounds have been recorded. Training, test and optimisation of design parameters can be performed by partitioning this dataset into training and test subsets, a standard practice in machine learning that is further discussed below. To obtain a fair evaluation reflecting the conditions of a real-world application where sounds and labels are unknown to the algorithms, the methods submitted to the \Acronym{DCASE} challenge were tested on a private dataset.
\subsubsection{Cross-validation} Recall from Figure \ref{fig:fra} that statistical models are learned from elements of the training data that belong to different classes, and therefore depend on the particular signals available for training. This represents a general problem of statistical inference occurring every time models are learned using a limited set of data, and is associated with a sampling error or bias. For example, to learn a statistical model of the sounds produced in the office environment, we would ideally need complete and continuous historical recordings from every office in the world. By only analysing data recorded from one or several offices we are bound to learn models that are biased towards the sounds present within the available signals. However, if the training data are rich enough to include sounds produced in most office environments, and if these sounds are effectively modelled, then the sampling bias can be bounded, and models can statistically infer general properties of office environments from an incomplete set of measurements.
Cross-validation is employed to minimise the sampling bias by optimising the use of a set of available data. The collection of labelled recordings is partitioned into different subsets for training and testing so that all the samples are used in the test phase. Different partition methods have been proposed in the literature for this purpose \cite{Bishop2007Pa}. To evaluate the algorithms submitted to the \Acronym{DCASE} challenge we employed a so-called stratified $5$-fold cross-validation of the private dataset. From $100$ available recordings, five independent classifications are performed, so that each run contains $80$ training recordings and $20$ test recordings. The partitions are designed so that the five test subsets are disjoint, thus allowing to perform the classification of each of the $100$ signals in the test phases. In addition, the proportion of signals belonging to different classes is kept constant in each training and test subset ($8$ signals per class in the former and $2$ signals per class in the latter) to avoid class biases during the statistical learning.
\subsubsection{Performance metrics} Performance metrics were calculated from each classification obtained using the training and test subsets, yielding $5$ results for each algorithm. Let $\acc$ be the set of correctly classified samples. The classification \emph{accuracy} is defined as the proportion of correctly classified sounds relative to the total number of test samples. The \emph{confusion matrix} is a $\nUniCat\times\nUniCat$ matrix whose $(i,j)$-th element indicates the number of elements belonging to the $i$-th class that have been classified as belonging to the $j$-th class. In a problem with $\nUniCat=10$ different classes, chance classification has an accuracy of $0.1$ and a perfect classifier as an accuracy of $1$. The confusion matrix of a perfect classifier is a diagonal matrix whose $(i,i)$-th elements correspond to the number of samples belonging to the $i$-th class.
\subsection{Results}
\begin{figure}[!tb]
\begin{center}
\includegraphics[width=.8\textwidth]{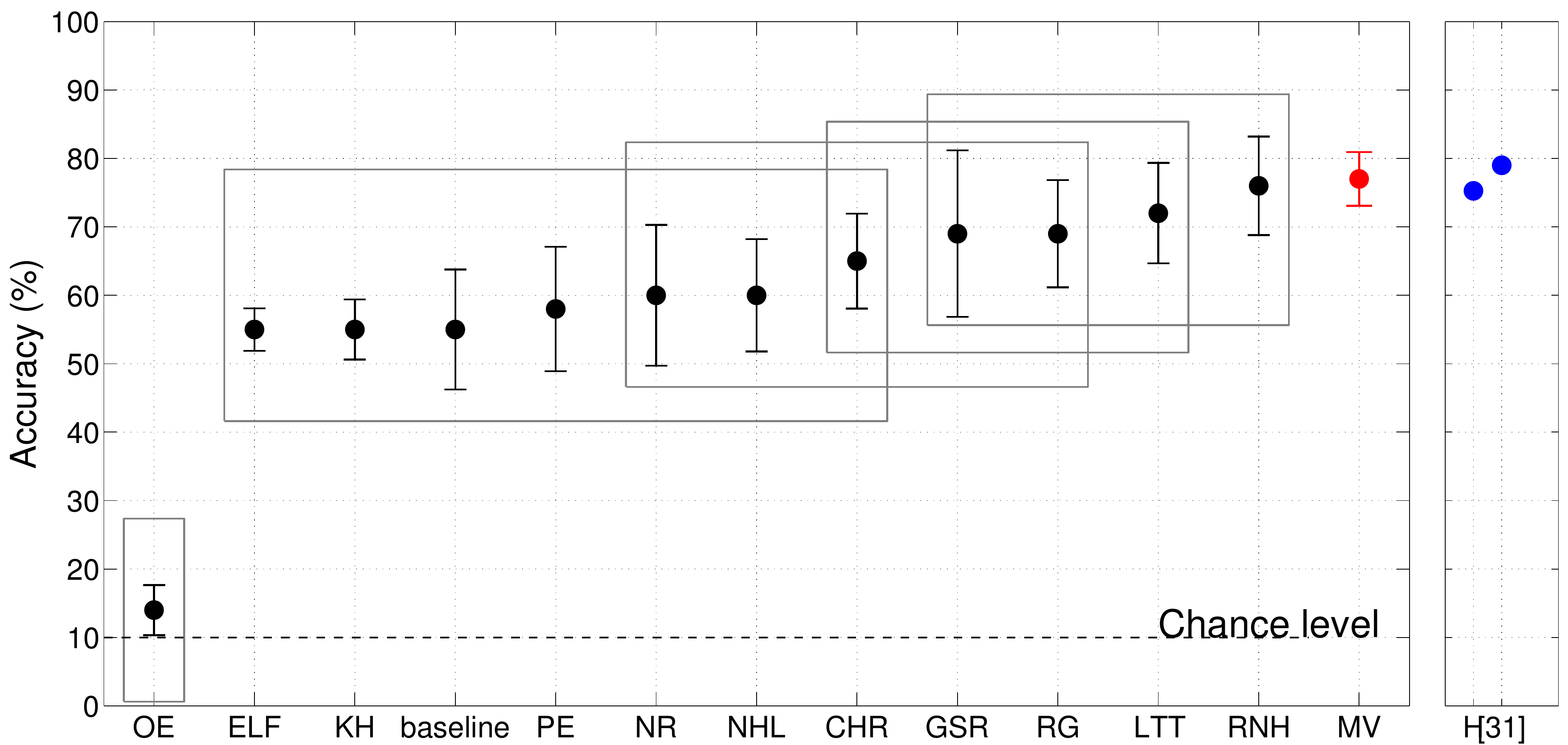}
\caption{\label{fig:ranking}Mean values and confidence intervals of the accuracy of methods for \Acronym{ASC} evaluated on the \Acronym{DCASE} private dataset using stratified 5-fold cross-validation. The boxes enclose methods that cannot be judged to perform differently with a significance level of $95\%$. Please see Table \ref{tab:dcase} for the definition of the algorithms' acronyms. \textcolor{red}{MV is a majority vote classifier which assigns to an audio recording the label that is most commonly returned by the other methods. `H' indicates the median human accuracy, as obtained through the test described in Section \ref{sec:hp}, while `[31]' refers to the human accuracy obtained by Krijnders and Holt. Note that the confidence intervals displayed for the algorithmic results are not directly comparable to the variations in human performance, and hence only the median human performance is depicted. See Figure \ref{fig:hacc} for more details on the distribution of human accuracies.}}
\end{center}
\end{figure}
Figure \ref{fig:ranking} depicts the results for the algorithms submitted to the \Acronym{DCASE} challenge (see Table \ref{tab:dcase} for the acronyms of the methods). The central dots are the percentage accuracies of each technique calculated by averaging the results obtained from the $5$ folds, and the bars are the relative confidence intervals. These intervals are defined by assuming that the accuracy value obtained from each fold is a realisation of a Gaussian process whose expectation is the true value of the overall accuracy (that is, the value that we would be able to measure if we evaluated an infinite number of folds). The total length of each bar is the magnitude of a symmetric confidence interval computed as the product of the $95\%$ quantile of a standard normal distribution $q^{0.95}_{\Normal(0,1)}\approx 3.92$ and the standard error of the accuracy (that is, the ratio between the standard deviation of the accuracies of the folds and the square root of the number of folds $\sigma/\sqrt{5}$). Under the Gaussian assumption, confidence intervals are interpreted as covering with $95\%$ probability the true value of the expectation of the accuracy. 
  
From analysing the plot we can observe that the baseline algorithm achieves a mean accuracy of $55\%$, and a group of other methods obtain a similar result in the range between $55\%$ and $65\%$. Four algorithms (\Acronym{GSR}, \Acronym{RG}, \Acronym{LTT} and \Acronym{RNH}) approach or exceed a mean accuracy of $70\%$. \Acronym{OE} performs relatively close to chance level and significantly worse than all the other methods. The boxes displaying the results of the paired tests explained in Section \ref{sec:rank} indicate that a number of systems performed significantly better than baseline.

\begin{figure}[!tb]
\begin{center}
\includegraphics[width=\textwidth]{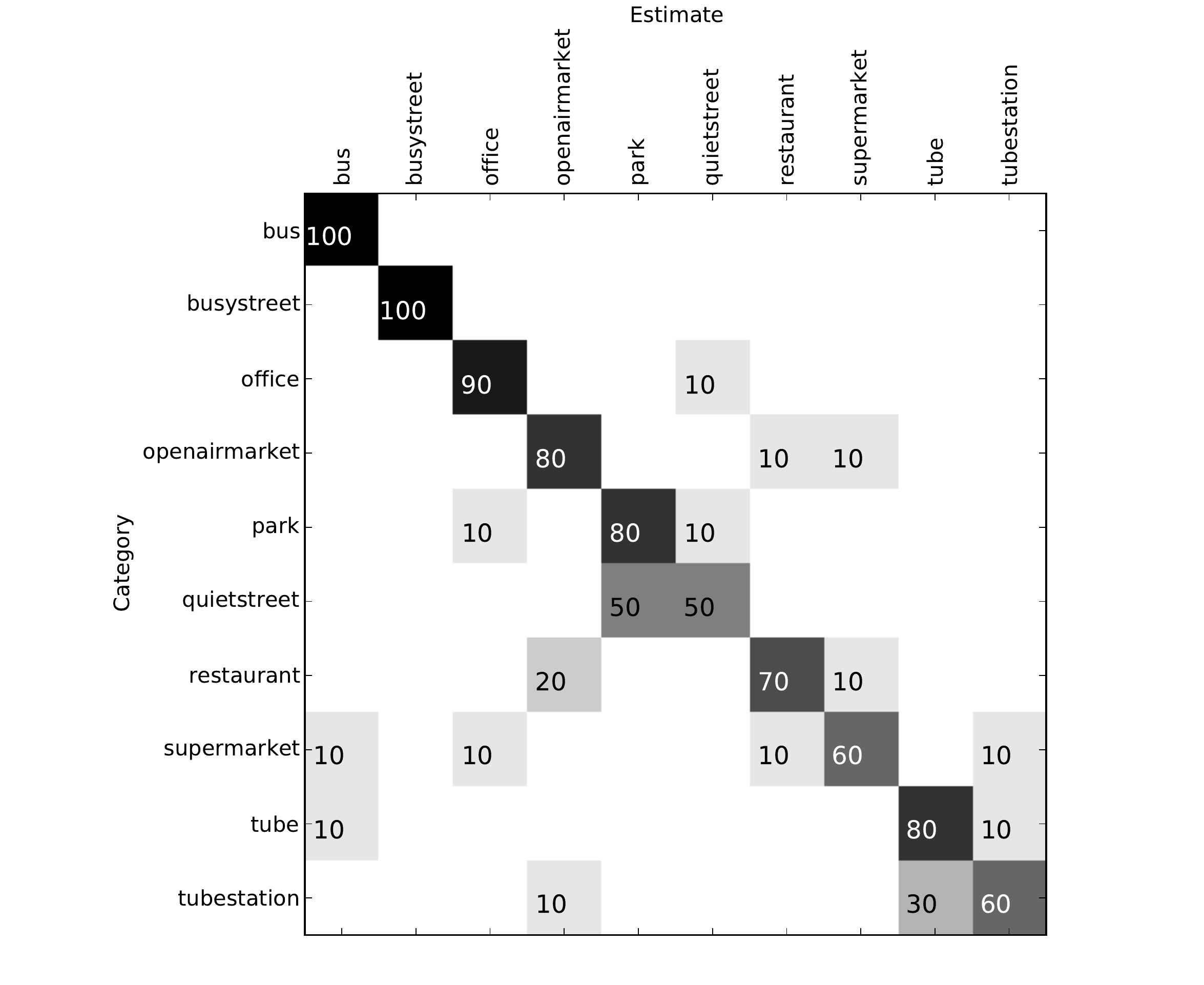}
\caption{\label{fig:aconmat}Confusion matrix of \Acronym{MV} algorithmic classification results.}
\end{center}
\end{figure}

Finally, the method \Acronym{MV} indicated in red refers to a majority vote classifier whose output for each test file is the most common category assigned by all other methods. The mean accuracy obtained with this meta-heuristic out-performs all the other techniques, indicating a certain degree of independence between the classification errors committed by the algorithms. \textcolor{red}{In other words, for almost $80\%$ of soundscapes some algorithms make a correct decision, and the algorithms that make an incorrect classification do not all agree on one particular incorrect label. This allows to combine the decisions into a relatively robust meta-classifier}. On the other hand, the performance obtained using \Acronym{MV} is still far from perfect, suggesting that a number of acoustic scenes are misclassified by most algorithms. \textcolor{red}{Indeed, this can be confirmed by analysing the confusion matrix of the \Acronym{MV} solution. As we can see in Figure \ref{fig:aconmat}, the class pairs (`park',`quietstreet') and (`tube',`tubestation') are commonly misclassified by the majority of the algorithms.} 

To investigate the poor performance of the method \Acronym{OE}, we considered the results obtained on the public \Acronym{DCASE} dataset, which are not detailed here for the sake of conciseness. \Acronym{OE} obtained the highest classification accuracy of all methods, suggesting that it over-fitted the training data by learning models that could not generalise to the test signals.

\subsection{Ranking of algorithms}\label{sec:rank}
The \Acronym{ASC} performance has been evaluated by computing statistics among different cross-validation folds. However, all the submitted methods have been tested on every file of the same held-back dataset, and this allows us to compare their accuracy on a file-by-file basis. Recall that $\sig_{\iTes}$ indicates a signal in the test set. A binary variable $\BernoulliVar_{\iTes}$ can be assigned to each signal and defined so that it takes the value $1$ if the file has been correctly classified and $0$ if it has been misclassified. Each $\BernoulliVar_{\iTes}$ can be thus interpreted as a realisation of a Bernoulli random process whose average is the mean accuracy of the classifier. 

Given two classifiers $\Classifier_{1}$, $\Classifier_{2}$, and the  corresponding variables $\BernoulliVar_{\Classifier_{1},{\iTes}}$,$\BernoulliVar_{\Classifier_{2},{\iTes}}$, a third random variable $\SignVar_{\iTes} = \BernoulliVar_{\Classifier_{1},{\iTes}}- \BernoulliVar_{\Classifier_{2},{\iTes}}$ assumes values in the set $\curlyb{-1,0,+1}$ and indicates the difference in the correct or incorrect classification of $\sig_{\iTes}$ by the two classifiers (that is, $\SignVar=-1$ implies that $\Classifier_{1}$ has misclassified $\sig$ and $\Classifier_{2}$ has correctly classified it; $\SignVar=0$ means that the two methods return equivalently correct or incorrect decisions, and $ \SignVar=1$ implies that $\Classifier_{1}$ has correctly classified $\sig$ and $\Classifier_{2}$ has misclassified it). A \emph{sign test} \cite{Gibbons2010No} can be performed to test the hypothesis that the expected value of $\SignVar$ is equal to zero. This is equivalent to performing a paired test evaluating the hypothesis that the performance of the two classifiers $\Classifier_{1}$ and $\Classifier_{2}$ is the same. Hence, being able to reject this hypothesis at a fixed probability level provides a method to rank the algorithms.

The grey boxes in Figure \ref{fig:ranking} represent groups of methods whose accuracy is not significantly different when tested on the \Acronym{DCASE} dataset, according to the sign tests ranking criterion evaluated between pairs of different methods. Methods enclosed in the same box cannot be judged to perform better or worse according to the chosen significance level. Starting with the least accurate algorithms, we can observe that the performance of \Acronym{OE} is significantly different compared with all the other techniques. Then a clusters of methods ranging from \Acronym{ELF} to \Acronym{CHR} do not perform significantly differently from the baseline. \Acronym{GSR} and \Acronym{RG} can be said to have significantly higher accuracy if compared to the baseline method, but not if compared to \Acronym{NR}, \Acronym{NHL} or \Acronym{CHR}. Finally \Acronym{RNH} is not significantly more accurate than \Acronym{GSR}, \Acronym{RG} and \Acronym{LTT}, but outperforms all the remaining methods. Note that we do not include the results of the majority vote meta-heuristic in the ranking, as a paired sign test assumes the variables $\BernoulliVar_{\Classifier_{1},\iTes}, \BernoulliVar_{\Classifier_{2},\iTes}$ to be statistically independent, and this assumption is violated in the case of \Acronym{MV}.


\subsection{Distribution of algorithmic soundscapes classification accuracies}\label{sec:asca}


\begin{figure}[!tb]
\begin{center}
\includegraphics[width=\textwidth]{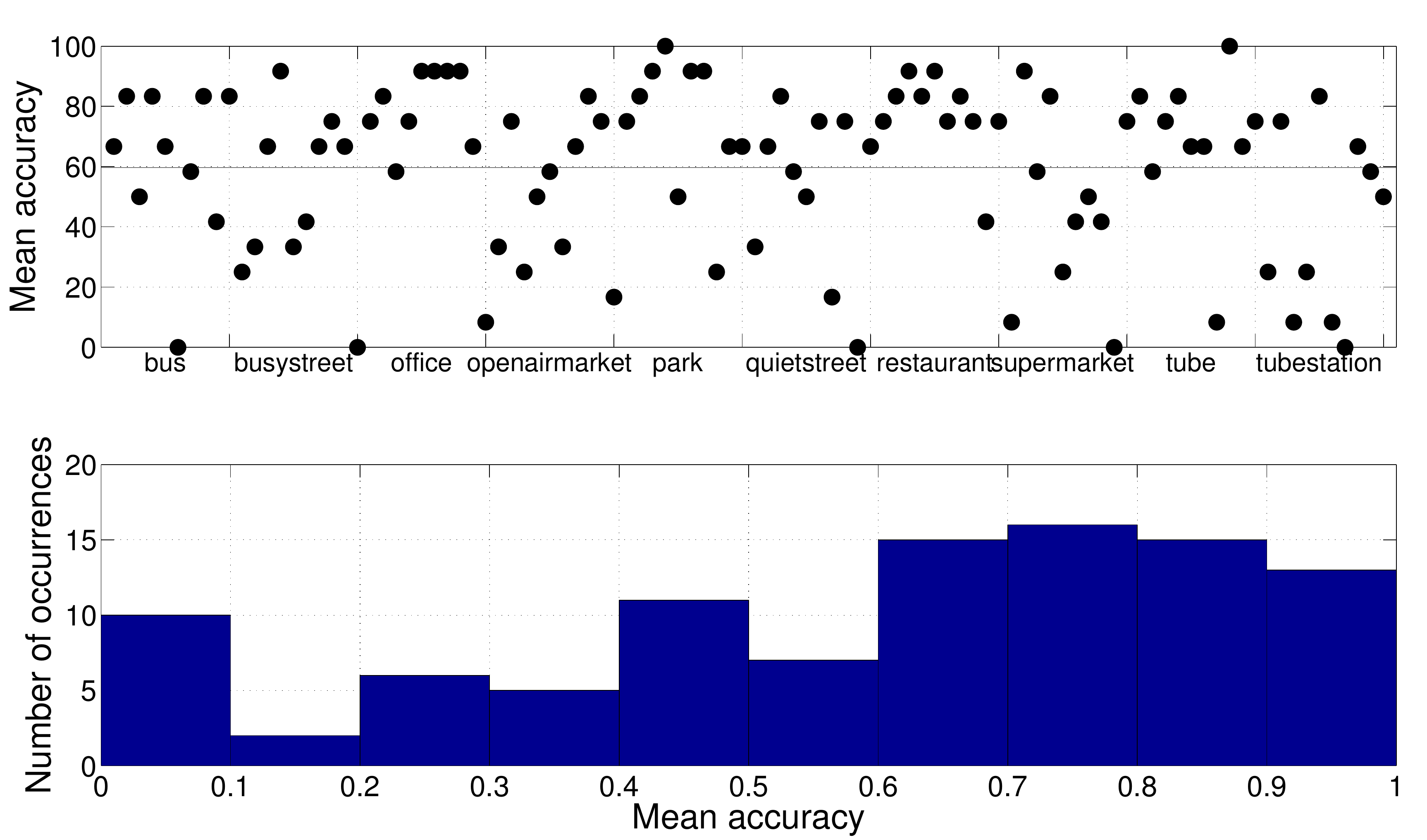}
\caption{\label{fig:sca}Distribution of algorithmic soundscapes classification accuracies. The solid line in the upper plot represents the average accuracy calculated from all the acoustic scenes. \textcolor{red}{The bottom plot depicts the histogram of mean accuracies resulting from the classification of all $100$ soundscapes, highlighting in the left tail that ten soundscapes correctly classified by at most only 10\% of the algorithms.}}
\end{center}
\end{figure}
Further analysis of the classification results can be carried out to understand whether there are individual soundscape recordings in the \Acronym{DCASE} dataset that are classified more accurately than others. After evaluating each method with a $5$-fold cross-validation, every signal $\sig_{\iTes}$ is classified by all the algorithms, resulting in a total of $12$ estimated categories. Figure \ref{fig:sca} shows a scatter-plot of the mean classification accuracy obtained for each file, and a histogram of the relative distribution. We can observe that some acoustic scenes belonging to the categories `bus', `busy-street', `quietstreet' and `tubestation' are never correctly classified (those at $0\%$). In general, the classification accuracy among soundscapes belonging to the same category greatly varies, with the exception of the classes `office' and `restaurant' that might contain distinctive events or sound characteristics resulting in more consistent classification accuracies. 

\subsection{Pairwise similarity of algorithms decisions}
\label{sec:mds}

\begin{figure}[!tb]
\begin{center}
\includegraphics[width=.9\textwidth]{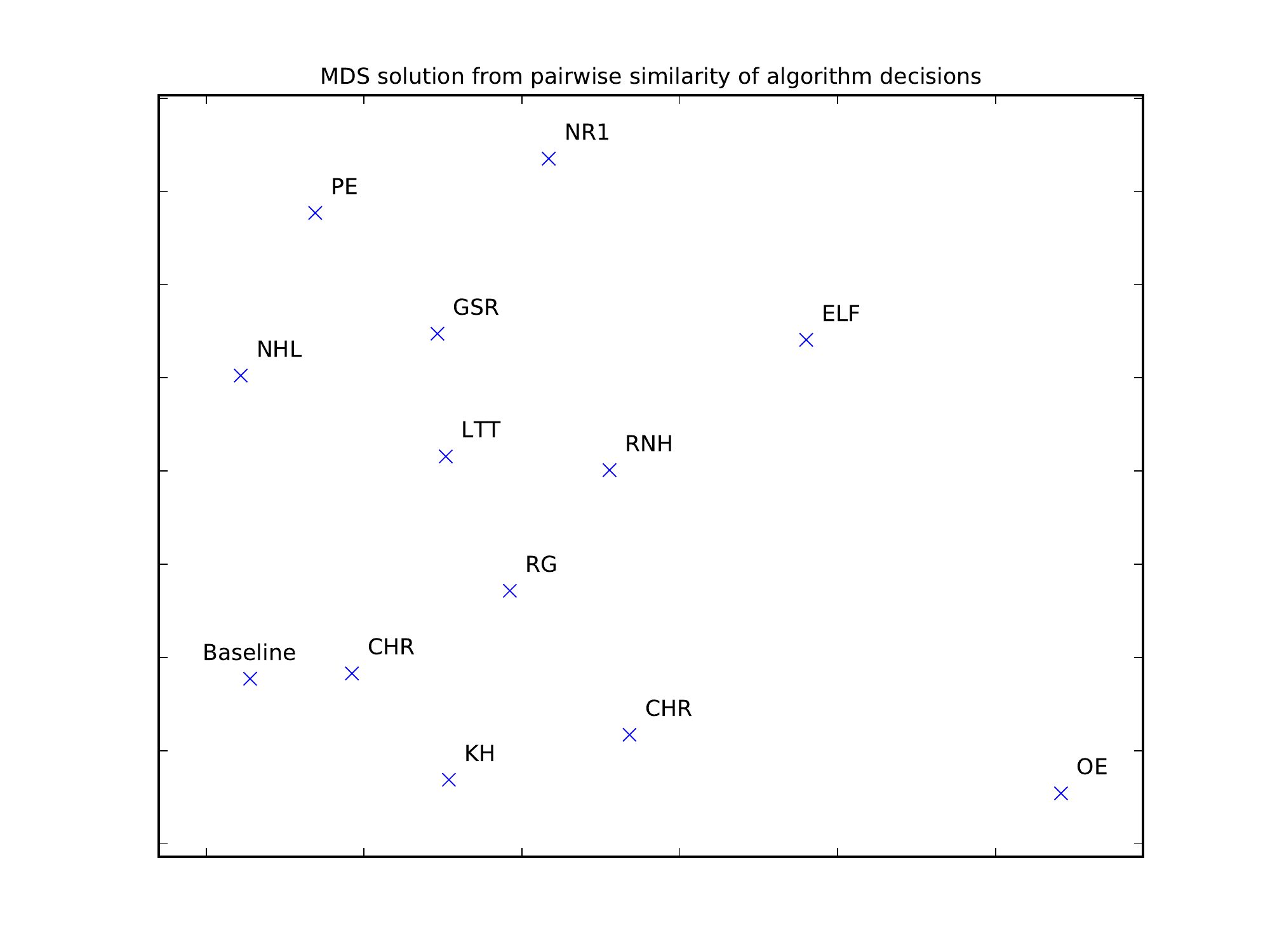}
\caption{\label{fig:plot_dcase_scenerec_mds}Multidimensional scaling solution (two-dimensional) derived from the pairwise similarities between algorithm labelling decisions. Algorithms which make similar (mis)classifications will tend to appear close to one another. See Section \ref{sec:mds} for details. %
}
\end{center}
\end{figure}

While the results in Figure \ref{fig:ranking} demonstrate the overall accuracy achieved by algorithms,
they do not show which algorithms tend to make the same decisions as each other.
For example, if two algorithms use a very similar method, we
would expect them to make a similar pattern of mistakes.
We can explore this aspect of the algorithms by comparing their decisions pairwise against one another,
and using the number of disagreements as a distance measure.
We can then visualise this using multidimensional scaling (MDS) to project the points into a low-dimensional space which approximately
honours the distance values \cite[chapter 10]{Duda:2000}.

Results of MDS are shown in Figure \ref{fig:plot_dcase_scenerec_mds}.
\textcolor{red}{We tested multiple dimensionalities and found that 2D (as shown) yielded a sufficiently
low stress to be suitably representative.
The OE submission is placed in a corner of the plot,
at some distance from the other algorithms;
that submission achieved low scores on the private testing data.
As a whole, the plot does not appear to cluster together methods by feature type,
as MFCC and non-MFCC approaches are interspersed, as are SVM and non-SVM approaches.}

\section{Human listening test}\label{sec:hp}
In order to determine a human benchmark for the algorithmic results on \Acronym{ASC}, we have designed a \textcolor{red}{crowdsourced} online listening test in which participants were asked to classify the public \Acronym{DCASE} dataset  by listening to the audio signals and choosing the environment in which each signal has been recorded from the $10$ categories `bus',`busy-street',`office',`openairmarket',`park',`quiet-street',`restaurant',`supermarket',`tube' and `tubestation'.

In designing the listening experiment we chose not to divide the classification into training and test phases because we were interested in evaluating how well humans can recognise the acoustic environments basing their judgement on nothing other than their personal experience. Participants were not presented with labelled training sounds prior to the test, nor were they told their performance during the test. 

To maximise the number of people taking the test, we have allowed each participant to classify as many acoustic scenes as he or she liked, while  randomising the order in which audio samples appeared in the test to ensure that each file had the same probability to be classified. To avoid potential biases, people who were likely to have worked with the data, and thus likely to know the class labels in advance, did not take the test.

\subsection{Human accuracy}
\begin{figure}[!tb]
\begin{center}
\includegraphics[width=.8\textwidth]{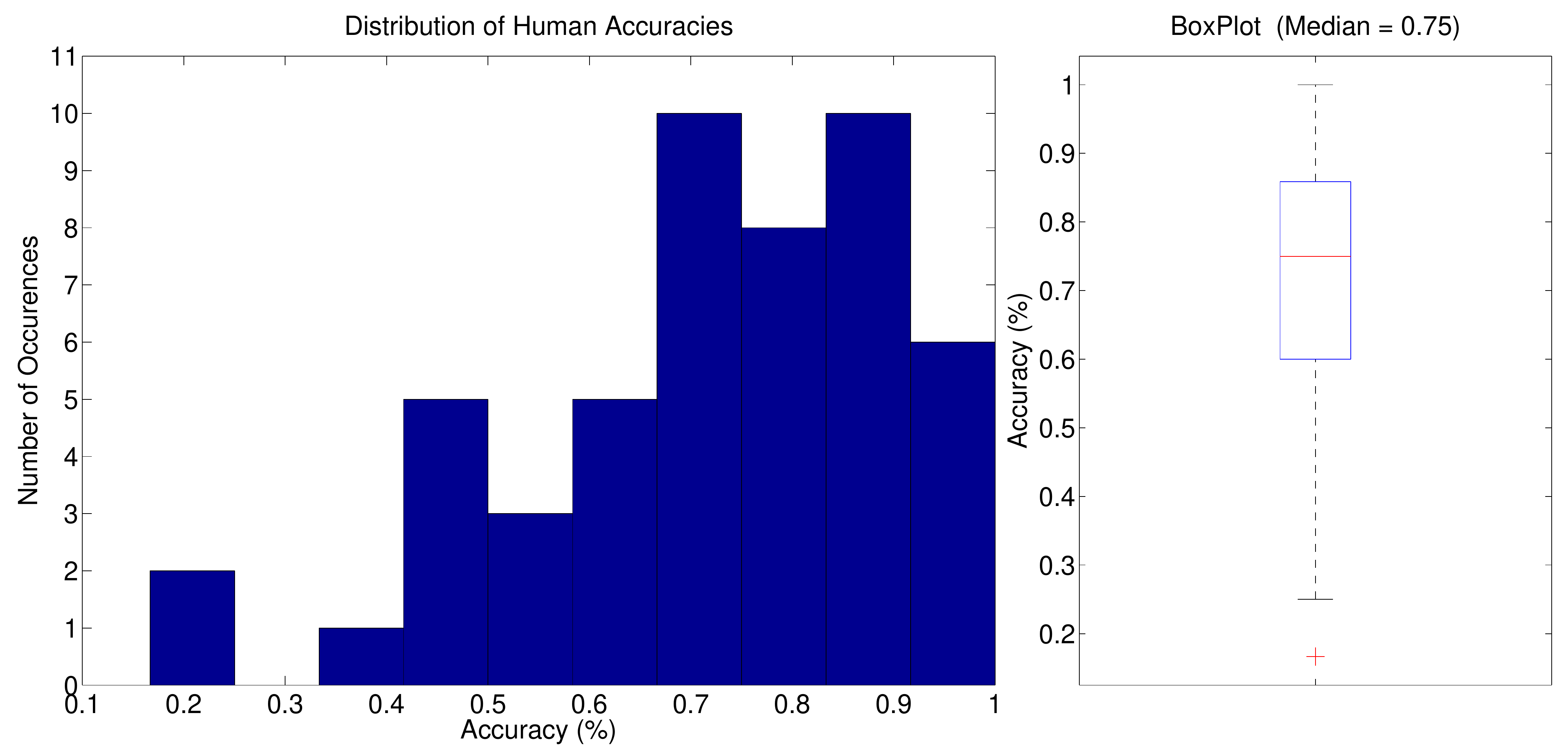}
\caption{\label{fig:hacc}Distribution of human soundscape classification accuracies.}
\end{center}
\end{figure}

A total of $50$ participants took part in the test. \textcolor{red}{Their most common age was between 25 and 34 years old, while the most common listening device employed during the test was ``high quality headphones''}. Special care was taken to remove ``test'' cases or invalid attempts from the sample. This included participants clearly labelled as ``test'' in the metadata, and participants who only attempted to label only $1-2$ soundscapes, and most of whom achieved scores as low as $0\%$ that points to outliers with a clear lack of motivation. Figure \ref{fig:hacc} shows that the mean accuracy among all participants was $72\%$, and the distribution of accuracies reveals that most people scored between $60\%$ and $100\%$, with two outlier whose accuracy was as low as $20\%$. Since the distribution of accuracies is not symmetric, we show a box plot summarising its statistics instead of reporting confidence intervals for the mean accuracy. The median value of the participants' accuracy was $75\%$, the first and third quartiles are located at around $60\%$ and $85\%$, while the $95\%$ of values lie between around $45\%$ and $100\%$. Note that, although we decided to include the results from all the participants in the study who classified at least a few soundscapes, the most extreme points (corresponding to individuals who obtained accuracies of about $25\%$ and $100\%$ respectively) only include classifications performed on less than $10$ acoustic scenes. Removing from the results participants who achieved about $25\%$ accuracy would result in a mean of $74\%$ a lot closer to the median value. \textcolor{red}{In a more controlled listening test, Krijnders and Holt \cite{Krijnders:2013} engaged $37$ participants, with each participant asked to listen to $50$ public DCASE soundscapes and select one of the $10$ categories. The participants were required to listen for the entire duration of the recordings, and use the same listening device. They obtained a mean accuracy of $79\%$, which is in the same area as the results of our crowdsourced study ($75\%$).}

\subsubsection{Cumulative accuracy}
During the test, we asked the participants to indicate their age and the device they used to listen to the audio signals, but we did not observe correlation between these variables and the classification accuracy. We did observe a correlation between the number of classified samples and the overall classification accuracy. People who listened to and categorised most or all of the $100$ total samples tended to score better than individuals who only classified a few sounds. To assess whether this occurred because participants learned how to better classify the sounds as they progressed in the test, we computed for each individual the cumulative accuracy $\cumAcc(\iTim)$, that is defined as the ratio between the number of correctly classified samples and the total number of classified samples at times $\iTim=1,\dots,\nTes$:
\begin{equation}
	\cumAcc(\iTim) = \frac{\abs{\acc(\iTim)}}{\iTim}.
\end{equation}
A positive value of the discrete first time derivative of this function $\cumAcc'(\iTim) = \cumAcc(\iTim)-\cumAcc(\iTim-1)$ would indicate that there is an improvement in the cumulative classification accuracy as time progresses. Therefore, we can study the distribution of $\cumAcc'(\iTim)$ to assess the hypothesis that participants have been implicitly training an internal model of the classes as they performed the test. The average of the function $\cumAcc'(\iTim)$ calculated for all the participants results to be $-0.0028$. A right-tailed t-test rejected with $95\%$ probability that the expectation of $\cumAcc'(\iTim)$ is greater than zero, and a left-tailed t-test failed to reject with the same probability the expectation is smaller that zero, indicating that participants did not improve their accuracy as they progressed through the test. This is a positive finding, as the listening test was designed to avoid training from the exposure to the soundscapes. Having rejected the learning hypothesis, we are left with a selection bias explanation: we believe that people who classified more sounds were simply better able or more motivated to do the test than individuals who found the questions difficult or tedious and did not perform as well.

\subsection{Scenes class confusion matrix}
\begin{figure}[!tb]
\begin{center}
\includegraphics[width=\textwidth]{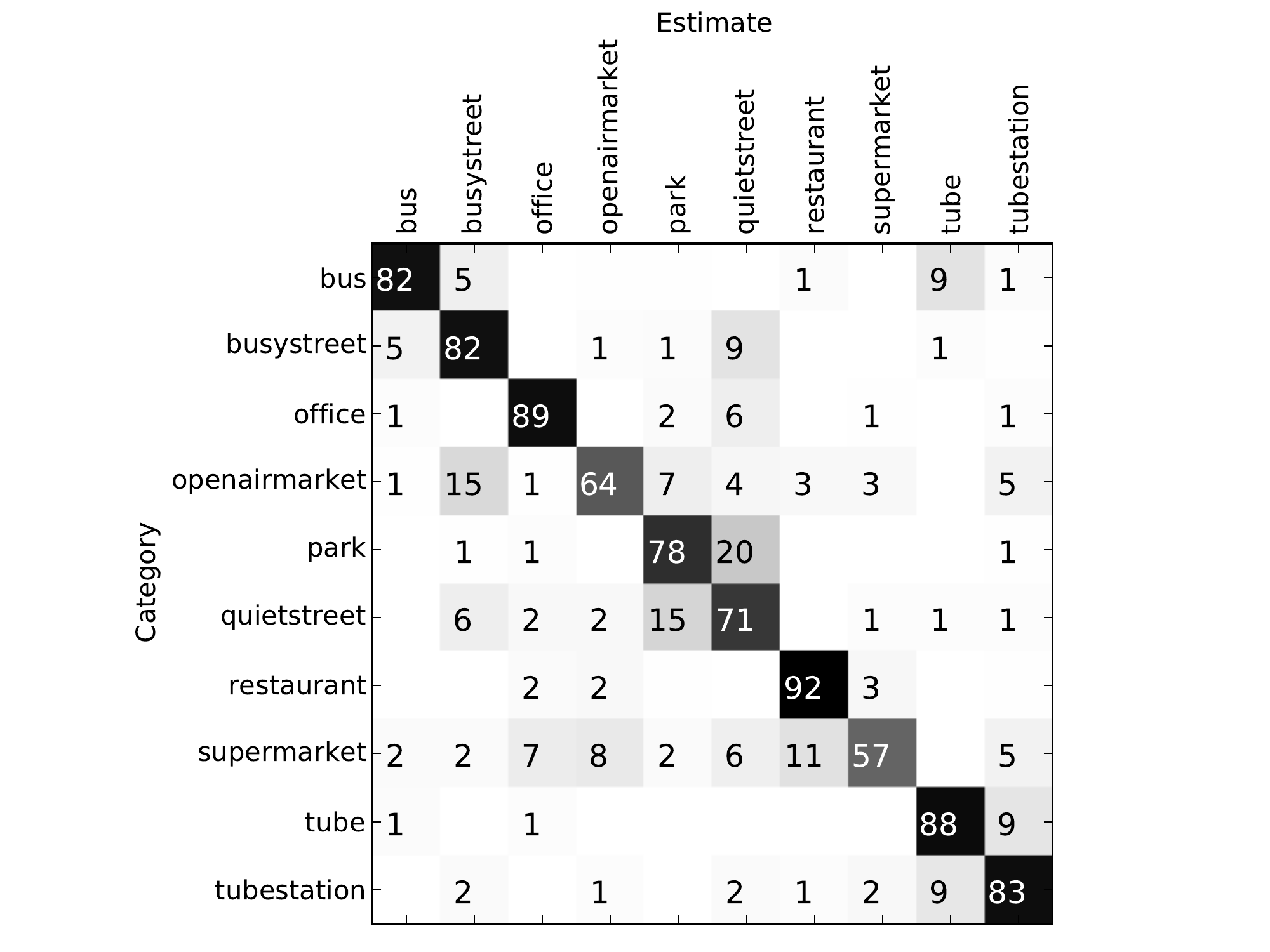}
\caption{\label{fig:hconmat}Confusion matrix of human classification results. Note that the rows of the confusion matrix might not add to $100\%$ due to the rounding of percentages}
\end{center}
\end{figure}
Further insight about the human classification results can be obtained by analysing the overall confusion matrix of the listening test. Figure \ref{fig:hconmat} shows that `supermarket' and `openairmarket' are the most commonly misclassified categories whose samples have been estimated as belonging to various other classes. In addition, there are some common misclassifications between the classes `park' and `quietstreet', and (to a minor extent) between the classes `tube' and `tubestation'.

\subsection{Distribution of human soundscapes classification accuracies}
\begin{figure}[!tb]
\begin{center}
\includegraphics[width=\textwidth]{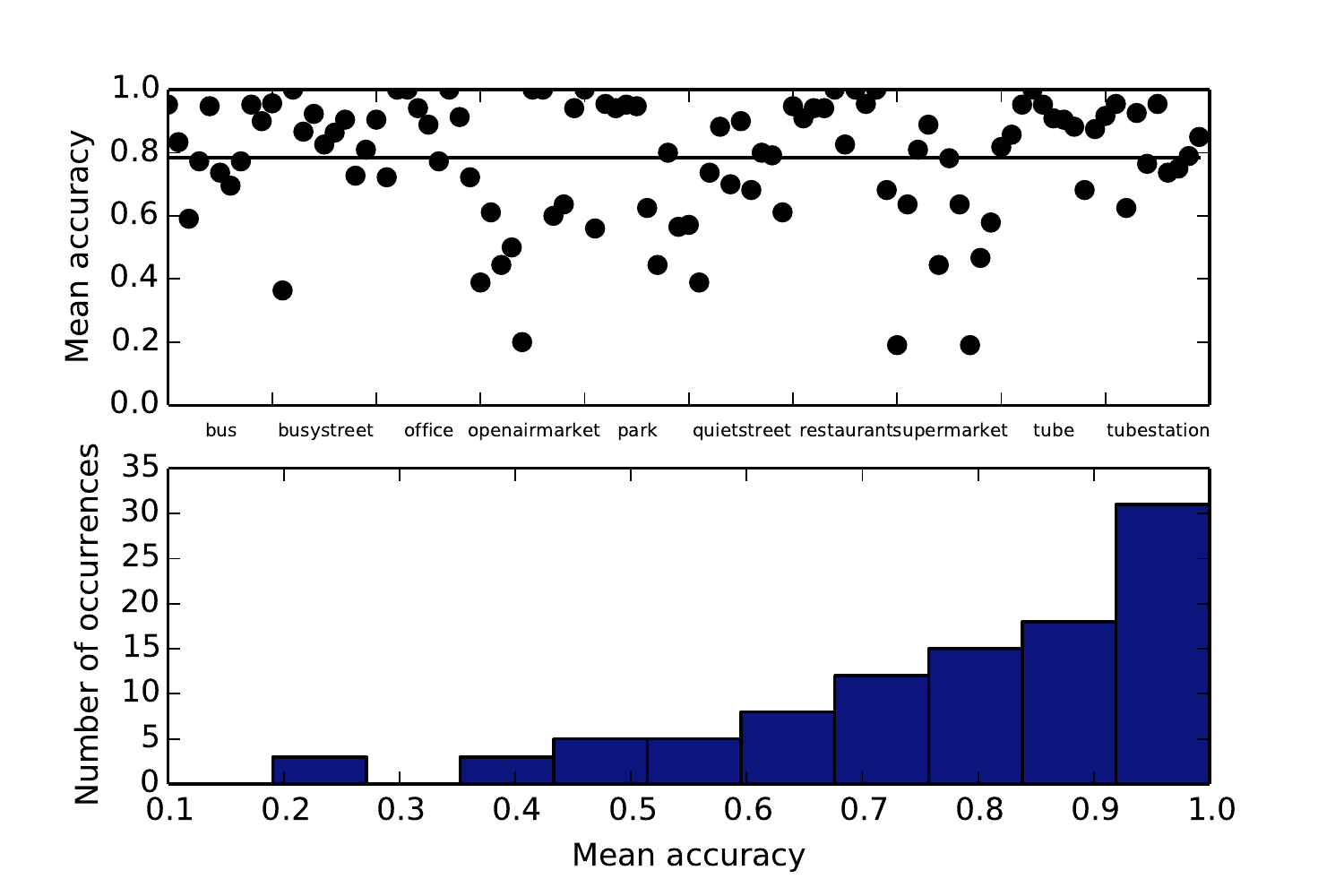}
\caption{\label{fig:lfa}Distribution of human soundscapes classification accuracies. The solid line in the upper plot represents the average accuracy.}
\end{center}
\end{figure}
To assess if some soundscapes were classified more accurately than others, we conducted a similar analysis for the human performance benchmark to the one described in Section \ref{sec:asca}. Figure \ref{fig:lfa} depicts the mean accuracy of classification of the $100$ soundscapes in the public \Acronym{DCASE} dataset, and a histogram of the relative distribution.  The public and private portions of the \Acronym{DCASE} dataset are disjoint subsets of the group of recordings produced for the challenge, therefore a paired comparison of the accuracies in Figures \ref{fig:sca} and \ref{fig:lfa} cannot be carried out. Nonetheless, it is informative to compare the trends between the two analysis: it appears that the mean performance for the human classification approaches $80\%$ as opposed to a value around $55\%$ achieved on average by the algorithms. In addition, the distribution of the mean accuracy in the case of a human classification appears more regular, with most soundscapes that are correctly classified most of the times, with only a few outlier scenes whose classification accuracy is below $30\%$.

\section{Discussion}\label{sec:dis}
By interpreting sophisticated algorithms in terms of a general framework, we have offered a tutorial that uncovers the most important factors to take into account when tackling a difficult machine learning task such as the classification of soundscapes. Inevitably, every abstraction or generalisation is carried out at the expense of omissions in the description of the implementation details of each method. Nonetheless, we think that valuable insights can be gained by analysing the classification results in light of the framework proposed in Section \ref{sec:fra}.
\subsection{Algorithms from the \Acronym{DCASE} challenge}\label{sec:meta}
A first trend regarding the choice of statistical learning function $\lea$ can be inferred by analysing the algorithms submitted for the \Acronym{DCASE} challenge summarised in Table \ref{tab:asc}. All but one method (\Acronym{ELF}) use discriminative learning to map features extracted from the audio signals $\sig_{\iSig}$ to class labels $\cat_{\iSig}$. Moreover, most of the algorithms whose mean accuracy is greater or equal than what achieved by the baseline method employ \Acronym{SVM}. All techniques that perform significantly better than the baseline except \Acronym{LTT} employ a combination of generative and discriminative learning by training an \Acronym{SVM} classifier using parameters of models $\model_{\iSig}$ learned from individual audio scenes. This suggests that models learned from single audio scenes offer an appropriate tradeoff between discrimination and generalisation. On one hand audio signals recorded in the same environment are analysed by learning different statistical models that account for variations between one recording and the next. On the other hand, the parameters of these models occupy localised regions in a parameters space, so that classification boundaries can be learned to discriminate between signals recorded in different environments.

A closer analysis of some of the better scoring algorithms (\Acronym{GSR}, \Acronym{RG} and \Acronym{RNH}) reveals a further common design motivation. In different ways, all three methods attempt to model temporal relationships between features extracted from different portions of the signals. \Acronym{RNH} employes \Acronym{RQA} parameters to encode periodicities (or stationarity) of the \Acronym{MFCC} coefficients, \Acronym{RG} accounts for time-frequency structures in the audio signals by learning gradient histograms of images derived from their spectrograms, and finally \Acronym{GSR} computes linear regression coefficients of local features that encode general trends across a whole scene. This supports the intuitive observation that an \Acronym{ASC} method should take into consideration the time evolution of different acoustic events to model complex acoustic scenes.

A further observation derived from analysing Table \ref{tab:asc} is that among the methods that used classification trees in combination with a  tree bagger or a random forest algorithm, \Acronym{OE} achieved a poor classification performance, while \Acronym{LTT} reached the second best mean accuracy. This might suggest that meta-algorithms can be a valuable strategy, but may also be prone to over-fitting. 

Finally, a more exploratory remark regards the general use of the framework described in Section \ref{sec:fra}. Aucoutourier \cite{Aucouturier2004Im} studied the performance of a class of algorithms for audio timbre similarity which followed a method similar to the \Acronym{ASC} baseline. He reported the existence of a ``glass ceiling'' as more and more sophisticated algorithms failed to improve the performance obtained using a simple combination of \Acronym{MFCCs} and \Acronym{GMMs}. To a certain extent, the fact that $7$ out of $11$ \Acronym{ASC} methods did not significantly outperform our baseline might suggest a similar effect, and urges researchers to pursue alternative paradigms. Modelling temporal relationships as described above is one first step in this direction; and perhaps algorithms whose design motivations depart from the ones driving the development of the baseline, such as the normalised compression dissimilarity (\Acronym{OE}), might be worth additional investigation.

\subsection{Comparison of human and algorithmic results}
When designing the human listening test, we chose to present individuals with samples from the public \Acronym{DCASE} dataset to avoid distributing the held-back dataset that was produced to test the algorithms. In addition, we chose not to divide the human task into training and testing phases because we were interested in evaluating how people performed by only drawing from previous experience, and not from prior knowledge about the test set. The different experimental design choices between human and algorithmic experiments do not allow us to perform a statistically rigorous comparison of the classification performances. However, since the public and private \Acronym{DCASE} datasets are two parts of a unique session of recordings realised with the same equipment and in the same conditions, we still believe that qualitative comparisons are likely to reflect what the results would have been had we employed a different design strategy that allowed a direct comparison. More importantly, we believe that qualitative conclusions about how well algorithms can approach human capabilities are more interesting than rigorous significance tests on how humans can perform according to protocols (like the 5-fold stratified cross-validation) that are a clearly unnatural task.

Having specified the above disclaimer, several observations can be derived from comparing algorithmic and human classification results. Firstly, Figures \ref{fig:ranking} and \ref{fig:hacc} show that \Acronym{RNH} achieves a mean accuracy in the classification of soundscapes of the private \Acronym{DCASE} dataset that is similar to the median accuracy obtained by humans on the public \Acronym{DCASE} dataset. This strongly suggests that the best performing algorithm does achieve similar accuracy compared to a median human benchmark.

Secondly, the analysis of misclassified acoustic scenes summarised in Figures \ref{fig:sca} and \ref{fig:lfa} suggests that, by aggregating the results from all the individuals who took part in the listening test, all the acoustic scenes are correctly classified by at least some individuals, while there are scenes that are misclassified by all algorithms. This observation echoes the problem of \emph{hubs} encountered in music information retrieval, whereby certain songs are always misclassified by algorithms \cite{Radovanovic2010Hu}. Moreover, unlike for the algorithmic results, the distribution of human errors shows a gradual decrease in accuracy from the easiest to the most challenging soundscapes. This observation indicates that, in the aggregate, the knowledge acquired by humans through experience still results in a better classification of soundscapes that might be considered to be ambiguous or lacking in highly distinctive elements.

Finally, the comparison of the confusion matrices presented in Figure \ref{fig:aconmat} and Figure \ref{fig:hconmat} reveals that similar pairs of classes (`park' and `quietstreet', or `tube' and `tubestation') are commonly misclassified by both humans and algorithms. Given what we found about the misclassification of single acoustic scene, we do not infer from this observation that the algorithms are using techniques which emulate human audition. An alternative interpretation is rather that some groups of classes are inherently more ambiguous than others because they contain similar sound events. Even if both physical and semantic boundaries between environments can be inherently ambiguous, for the purpose of training a classifier the universe of soundscapes classes should be defined to be mutually exclusive and collectively exhaustive. In other words, it should include all the possible categories relevant to an \Acronym{ASC} application, while ensuring that every category is as distinct as possible from all the others.


\subsection{Further research}
Several themes that have not been considered in this work may be important depending on particular \Acronym{ASC} applications, and are suggested here for further research.
\subsubsection{Algorithm complexity} A first issue to be considered is the complexity of algorithms designed to learn and classify acoustic scenes. Given that mobile context-aware services are among the most relevant applications of \Acronym{ASC}, particular emphasis should be placed in designing methods that can be run with the limited processing power available to smartphones and tablets. The resources-intensive processing of training signals to learn statistical models for classification can be carried out off-line, but the operators $\tra$ and $\dec$ still need to be applied to unlabelled signals and, depending on the application, might need to be simple enough to allow real-time classification results.
\subsubsection{Continuous and user-assisted learning} Instead of assuming a fixed set of categories as done in most publications on \Acronym{ASC}, a system might be designed to be progressively trained to recognise different environments. In this case, a user should record soundscape examples that are used to train classification models (either on-line or off-line, using the recording device's own computational resources or uploading and processing the signals with remote cloud resources), and progressively add new categories to the system's \emph{memory} of soundscapes. Users could also assist the training by confirming or rejecting the category returned from querying each unlabelled signal, and thus refine the statistical models every time a new classification is performed. Such systems would inevitably require more intervention by the user, but would likely result to be more precise and relevant than totally automated systems. 
\subsubsection{Hierarchical classification} In this paper we have considered a set of categories whose elements are assumed to be mutually exclusive (that is, a soundscape can be classified as `bus' or `park' but not both). Alternatively, a hierarchical classification could be considered where certain categories are subsets or supersets of others. For example, a system might be designed to classify between `outdoor' and `indoor' environments, and then to distinguish between different subsets of the two general classes. In this context, different costs could be associated with different types of misclassification errors: for example, algorithms could be trained to be very accurate in discriminating between `outdoor' and `indoor', and less precise in distinguishing between an outdoor `park' and an outdoor `busy-street'.
\subsubsection{Acoustic scene detection} As a limit case of systems that employes non-uniform misclassification costs, algorithms might be designed to detect a particular environment and group all the other irrelevant categories into an `others' class. In this case, the system would essentially perform an acoustic scene detection rather than classification.
\subsubsection{Multi-modal learning} Another avenue of future research consists in fusing multi-modal information to improve the classification accuracy of \Acronym{ASC} systems. Video recordings, geo-location information, or temperature and humidity sensors are all examples of data that can be used in conjunction with audio signals to provide machines with context awareness.
\subsubsection{Event detection and scene classification} The combination of event detection algorithms and \Acronym{ASC} which has already been object of research endeavours \cite{Heittola2010Au,Chaudhuri2011Un} is likely to benefit from advances in both areas. Information regarding the events occurring in an acoustic scene could be combined with more traditional frame-based approaches to update the probability of categories as different events are detected. For example, while general spectral properties of a soundscape could be used to infer that a signal was likely to have been recorded in either a `park' or a `quiet street', detecting the event `car horn' would help disambiguate between the two. Furthermore, this Bayesian strategy employed to update the posterior probability of different classes could be used to handle transitions between different environments.
\subsubsection{Testing on different datasets}
Finally, datasets that contain sounds from different acoustic environments have been recently released. They include the Diverse Environments Multi-channel Acoustic Noise Database (\Acronym{DEMAND}) \cite{thiemann:DEMAND} and the Database of Annotated Real Environmental Sounds (\Acronym{DARES}) \cite{vandares}.

\textcolor{red}{\section{Conclusions}
In this article we have offered a tutorial in \Acronym{ASC} with a particular emphasis on computational algorithms designed to perform this task automatically. By introducing a framework for \Acronym{ASC}, we have analysed and compared methods proposed in the literature in terms of their modular components. We have then presented the results of the DCASE challenge, which set the state-of-the-art in computational \Acronym{ASC}, and compared the results obtained by algorithms with a baseline method and a human benchmark. On one hand, many of the submitted techniques failed to significantly outperform the baseline system, which was designed to be not optimised for this particular task. On the other hand, some methods significantly out-performed the baseline and approached an accuracy comparable to the human benchmark. Nonetheless, a more careful analysis of the human and algorithmic results highlighted that some acoustic scenes were misclassified by all algorithms, while all soundscapes were correctly classified by at least some individuals. This suggests that there is still scope for improvement before algorithms reach and surpass the human ability to make sense of their environment based on the sounds it produces.}

\section{Acknowledgments}
We would like to thank Dan Ellis, Toumas Virtanen, Jean-Julien Aucouturier, Mathieu Lagrange, Toni Heittola and the anonymous reviewers for having read and commented an early draft of this paper and on our submitted manuscript. Their insights and suggestions have substantially increased the value of this work. We also would like to thank the IEEE AASP Technical Committee for their support in the organisation of the \Acronym{DCASE} challenge.

\bibliographystyle{abbrv}
\bibliography{bibliography}

\appendix
\section{A baseline system for computational \Acronym{ASC}}\label{sse:a-b}
\subsection{MFCCs} Mel-frequency cepstral coefficients have been introduced in Section \ref{sse:ftm} and have been widely used as a feature for audio analysis. Let $\sig_{\iFra} \in \real^{\nFraDim}$ be a signal frame and $\abs{\Fourier{\sig}_{\iFra}}$ the absolute value of its Fourier transform. The coefficients corresponding to linearly-spaced frequency bins are mapped onto $\nMelBands$ Mel frequency bands to approximate the human perception of pitches (which can be approximately described as logarithmic, meaning that we are capable of a much better resolution at low frequencies than at high frequencies), resulting in $L\leq \nFraDim$ coefficients. The magnitude of the Mel coefficients is converted to a logarithmic scale, and the resulting vector is processed using a discrete cosine transform (\Acronym{DCT}). Finally,  the $\nFeaDim\leq \nMelBands$ first coefficients are selected and constitute the vector of features $\fea_{\iFra} = \tra(\sig_{\iFra})$. This last step essentially measures the frequency content of the log-magnitude of the spectrum of a signal, and therefore captures general properties of the spectral envelope. For example, periodic sounds which exhibit spectral peaks at multiples of a fundamental frequency are highly correlated with one or several cosine bases, encoding this information in the value of the corresponding \Acronym{MFCC} coefficients. The set of parameters $\parTra = \curlyb{\nFraDim,\nMelBands,\nFeaDim}$ includes frames dimension, number of Mel bands and number of \Acronym{DCT} coefficients which need to be defined when computing the \Acronym{MFCCs}. These parameters determine the dimensionality reduction introduced by the features extraction operator, and their choice is governed by the tradeoff between generalisation and discrimination that has been mentioned in Section \ref{sec:fra}.

\subsubsection{Statistical normalization}
To classify features extracted from signals belonging to different categories, it is important to evaluate relative differences between the values of feature vectors belonging to different classes, rather than differences between different coefficients within feature vectors extracted from the same signal. For this reason, during the training phase of the \Acronym{ASC} classification algorithm, statistical normalisation is performed as a standard feature processing aimed at avoiding offsets or scaling variations of any of the coefficients within feature vectors. This is accomplished by subtracting the global mean (computed from features extracted from the whole dataset) from each vector $\fea_{\iFra,\iSig}$, and by dividing each coefficient by the their global standard deviation. After the feature vectors have been normalised, the average and standard deviation of the coefficients $\feaCoeff_{\iFra,\iSig,\iFeaDim}$ are $0$ and $1$ respectively.

\subsection{GMMs} Gaussian mixture models (\Acronym{GMMs}) have been introduced in Section \ref{sec:sm}, and are used to infer global statistical properties of the features from local features vectors, which are interpreted as realisations of a generative stochastic process. Let $\Normal(\mean,\covar)$ be a multivariate normal distribution with mean $\mean\in\real^{\nFeaDim}$ and covariance matrix $\covar\in\real^{\nFeaDim\times\nFeaDim}$, and recall that the notation $\fea_{\iFra,\idxSet_{\iUniCat}}$ identifies features vectors extracted from training signals that belong to the $\iUniCat$-th category. Then every such vector is modelled as generated by the following distribution:
\begin{equation}\label{eq:gmm}
\fea_{\iFra,\idxSet_{\iUniCat}} \sim \prod_{\iCom=1}^{\nCom} \GaussWeight_{\iCom}\Normal(\mean_{\iCom},\covar_{\iCom})
\end{equation}
where $\nCom$ is a fixed number of components, and $\GaussWeight_{\iCom}$ is a latent variable expressing the probability that a particular observation is generated from the $\iCom$-th component.

The operator $\lea$ takes the collection of features $\fea_{\iFra,\idxSet_{\iUniCat}}$ and learns a global model for the $\iUniCat$-th class $\model_{\iUniCat} = \curlyb{\GaussWeight_{\iCom},\mean_{\iCom},\covar_{\iCom}}_{\iCom=1}^{\nCom}$ by   estimating the parameters of the Gaussian mixture distribution in Equation \eqref{eq:gmm}, which can be accomplished through an expectation-maximisation  (\Acronym{em}) algorithm \cite{Bishop2007Pa}. The only parameter to be set in this case is the number of Gaussian components $\nCom$ which rules a tradeoff between model accuracy and over-fitting. Indeed $\lea_{\nCom}$ must include a sufficient number of components to account for the fact that different events within a soundscape generate sounds with different spectral properties. However, as the number of components becomes too large, the model tends to fit spurious random variations in the training data, hindering the generalisation capabilities of the algorithm when confronted with an unlabelled sound.

\subsection{Maximum likelihood criterion} Once the \Acronym{GMMs} $\model_{\iUniCat}$ have been learned from the training data, features can be extracted from an unlabelled sound by applying the operator $\tra$. The new sequence of features $\fea_{\iFra,\new}$ is statistically normalised using the same mean and standard deviation values obtained from the training signals, and a likelihood measure $\dec$ is employed to evaluate which class is statistically most likely to generate the observed features, hence determining the sound classification. A set of coefficients $\crit_{\iUniCat}$ is computed by evaluating the log-likelihood of the observed data given the model:
\begin{equation}
\crit_{\iUniCat} = \prob(\fea_{\iFra,\new}|\model_{\iUniCat}) \propto \sum_{\iCom=1}^{\nCom} \GaussWeight_{\iCom} \Transpose{(\fea_{\iFra,\new}-\mean_{\iCom})}\covar_{\iCom}(\fea_{\iFra,\new}-\mean_{\iCom})
\end{equation}
and a category is picked based on the most likely model $\optimal{\cat_{\new}} = \Minimise{\iUniCat} \crit_{\iUniCat}$.

Note that the baseline system described in this section is an example of a bag-of-frames technique where the ordering of the sequence of features is irrelevant. Indeed, any random permutation of the sequences $\fea_{\iFra,\idxSet_{\iUniCat}}$ does not affect the computation of the \Acronym{GMM} parameters, and thus the classification of unlabelled signals.

\end{document}